\documentclass[11pt, letterpaper]{article}

\usepackage{epic}
\usepackage{eepic}
\usepackage{graphicx}
\usepackage{algorithm,algorithmic}
\usepackage{tikz}
\usepackage{xcolor,colortbl,ragged2e,rotating}
\usepackage{wrapfig}
\usepackage{amsmath,amsfonts,amssymb,amsthm}
\usepackage{listings}
\usepackage{spverbatim}
\usepackage{subcaption}
\usepackage{hyperref}
\usepackage{mathtools}
\usepackage{lineno}
\usepackage{float}
\usepackage{array}
\usepackage{longtable}
\usepackage{textcomp}
\usepackage{varioref}
\usepackage{xspace}
\usepackage{makeidx}
\usepackage{verbatim}
\usepackage{ifpdf}
\usepackage{tabularx}
\usepackage{keyval}
\usepackage{caption}
\usepackage{cite}
\usepackage{authblk}
\usepackage[left=2.00cm, right=2.00cm, top=2.00cm, bottom=2.00cm]{geometry}
\usepackage{varwidth}
\DeclareMathOperator{\sech}{sech}
\graphicspath{{Beam_graphics/}{Beam_tcache/}{Beam_gcache/}}
\DeclareGraphicsExtensions{.pdf,.eps,.ps,.png,.jpg,.jpeg}

\title{\bf{Hidden Asymptotic Symmetry in Long Elastic Beams on Softening Foundations}}

\author{Shrinidhi S. Pandurangi\textsuperscript {\dag}, Timothy J. Healey\textsuperscript {\ddag \dag} and Nicolas Triantafyllidis\textsuperscript {\S \P } \\ \relax \textsuperscript {\dag}Field of Theoretical and Applied Mechanics, Cornell University, Ithaca, NY, USA \\ \textsuperscript {\ddag }Department of Mathematics, Cornell University, Ithaca, NY, USA \\ \textsuperscript {\S }Laboratoire de Mécanique des Solides, C.N.R.S. UMR7649 \'Ecole Polytechnique, Palaiseau, France \\ \textsuperscript{\P}Aerospace Engineering Department \& Mechanical Engineering Department (emeritus), The University of Michigan, Ann Arbor, MI, USA}

\begin{document}

\maketitle

\begin{abstract}
Transverse wrinkles are known to appear in thin rectangular elastic sheets when stretched in the long direction. Numerically computed bifurcation diagrams for extremely thin, highly stretched films indicate entire orbits of wrinkling solutions, cf. Healey, et. al. [J. Nonlinear Sci., 23 (2013), pp.~777--805]. These correspond to arbitrary phase shifts of the wrinkled pattern in the transverse direction. While such behavior is normally associated with problems in the presence of a continuous symmetry group, an unloaded rectangular sheet possesses only a finite symmetry group. In order to understand this phenomenon, we consider a simpler problem more amenable to analysis -- a finite-length beam on a nonlinear softening foundation under axial compression.  We first obtain asymptotic results via amplitude equations, that are valid as a certain non-dimensional beam length becomes sufficiently large. We deduce that any two phase-shifts of a solution differ from one another by exponentially small terms in that length. We validate this observation with numerical computations, indicating the presence of solution orbits for sufficiently long beams. We refer to this as ``hidden asymptotic symmetry”.
\end{abstract}

\section{Introduction}\label{sec:Section_1}

We address a question in this work that was previously raised in the context of wrinkling in highly stretched thin elastic membranes \cite{Healey_Li_Cheng}. When a rectangular sheet, say, of length $L$ and width $W$, with $L>W$, is uniaxially stretched between rigid end grips in the longer direction, transverse wrinkles often emerge.  The phenomenon is well known, e.g., \cite{Friedl_Rammerstorfer_Fisher, Cerda_Ravi-Chandar_Mahadevan, Jacques_Potier-Ferry, Nayyar_Ravi-Chander_Huang}. The question in \cite{Healey_Li_Cheng} concerns an observed property of the transverse wrinkling patterns. Numerical computations were first carried out on half of the domain, of width $W/2$, for two distinct cases, assuming: (1) reflection symmetry leading to symmetric boundary conditions along the cut edge; (2) reflection-inversion symmetry leading to anti-symmetric boundary conditions there.  The two boundary value problems give rise to transverse wrinkled patterns that are (1) even and (2) odd, respectively, about the mid-plane. Remarkably the two resulting global bifurcation diagrams are identical (to within several significant digits commensurate with the accuracy of the computations), cf. \cite{Healey_Li_Cheng}. Additional computations were subsequently carried out employing the full rectangular domain. Transverse wrinkled patterns of arbitrary phase - neither symmetric nor anti-symmetric, in general - were obtained. Again, these yield precisely the same global bifurcation diagram, independent of the phase. Moreover, as the phase is increased (or decreased) an entire closed orbit of computed solutions, connecting the symmetric and anti-symmetric patterns, is observed. In other words, the system behaves as though it possesses a continuous symmetry group in the transverse direction.

Yet, the complete symmetry group of the rectangular reference configuration in $\mathbb{R}^3$ is finite, and its irreducible representations are all 1-dimensional. Consequently, the physical symmetry of the problem does not account for the observed degeneracy. The question is - what does? The wrinkling problem addressed in \cite{Healey_Li_Cheng} is difficult to analyze. First, there is no closed-form trivial (planar) solution. Worse, the domain has true corners; the solutions are not classical. Therefore, a rigorous bifurcation analysis, much less an analysis of the above-mentioned behavior, is apparently out of reach. We consider instead a simple model that is amenable to analysis, viz., an axially compressed, linear beam on a softening, nonlinear elastic foundation. We show that the model exhibits ``asymptotic" continuous symmetry for very long beams, i.e., when a certain non-dimensional length is sufficiently large. A precise comparison of this result with transverse membrane wrinkling is not possible, mostly due to the fact that the compressive stress distribution in the membrane is not uniform. However, an analogy does follow: The non-dimensional length of the beam approaching infinity is equivalent to the depth of the actual cross-section approaching zero, while the non-dimensional width of the membrane approaching infinity is equivalent to the actual membrane thickness approaching zero. 

Two-scale asymptotic approaches for beams resting on an elastic foundation or fluid supported beams are well known\cite{Lange_Newell, Potier-Ferry_1, Potier-Ferry_2, Hunt_Bolt_Thompson, Hunt_Wadee, Hunt_Wadee_Shiacolas, Audoly, Oshri_Brau_Diamant}. For infinite-length beams these correspond to a one-parameter family of asymptotic solutions that rapidly decay at $\pm \infty$, \cite{Diamant_Witten_1, Rivetti, Diamant_Witten_2, Rivetti_Neukirch}. An arbitrary phase shift acting on the ``fast" variable serves as the parameter, and an even solution connects to an odd one as the phase varies. Here we pursue the case of very long, simply-supported finite-length beams and obtain a novel, additive correction to the infinite-length solution. A non-zero phase shift of the corrected solution generally violates the boundary conditions. Nonetheless, the correction term is found to be an exponentially decaying function of the length of the beam. Accordingly, we deduce, for sufficiently long beams, that all phase-shifts acting on the fast variable yield equivalent solutions modulo exponentially small terms. We refer to this as {\it asymptotic symmetry}. Clearly the correction terms are not detectable by numerical methods for sufficiently long beams. We illustrate this via finite-element solutions of the boundary value problem, which show excellent agreement with the asymptotic solution. In the same way, we deduce that the energy difference between phase-shifted solutions is exponentially small. We first extend the work in \cite{Potier-Ferry_2}, giving complete post-critical analyses near the secondary bifurcation points via the amplitude equations. We compute global bifurcation diagrams, verifying that the corrected asymptotic solutions coincide with secondary bifurcations for very long beams of finite length.

The outline of the work is as follows. After briefly summarizing a non-dimensional formulation for the problem in Section \ref{sec:Section_2}, we take up bifurcation analyses in Section \ref{sec:Section_3}.  In Section \ref{sec:Section_3_1} we consider bifurcation of solutions from the compressed straight state. We fix the compressive loading parameter and prove the existence of bifurcating solutions that are even and odd, respectively, for lengths that are odd and even multiples of $\pi$, respectively. These nontrivial solutions can be identified with spatially periodic solutions. In Section \ref{sec:Section_3_2} we focus on secondary bifurcations from the principal path of odd-periodic solutions found in Section \ref{sec:Section_3_1} (secondary bifurcations from the even-periodic solutions of Section \ref{sec:Section_3_1} follow in the same manner). Following the lead of \cite{Potier-Ferry_1}, we seek 2-scale solutions and obtain the so-called amplitude equation. The primary branch of periodic solutions now appears as a constant (amplitude), from which we obtain local bifurcating solutions of the amplitude equations, representing secondary bifurcating solutions of the boundary value problem.  

In Section \ref{sec:Section_4} we analyze the amplitude equations, obtaining an asymptotic correction to the infinite-length solution for the very long, simply supported beams. The additive correction decays exponentially as a function of the length of the beam. In Section \ref{sec:Section_5} we present finite-element solutions for sufficiently long beams, demonstrating the accuracy of the asymptotic solutions and the numerical equivalence of all phase-shifted solutions, as described above. For sufficiently long beams, we compute an apparent closed orbit of solutions connecting odd to even solutions, i.e., the numerical model behaves as though it possesses a continuous symmetry group. We conclude with some final remarks in Section \ref{sec:Section_6}.

\section{Beam on a Nonlinear Elastic Foundation}\label{sec:Section_2}

We consider a linear elastic beam on a nonlinear elastic foundation, characterized by a softening cubic nonlinearity. The simply-supported beam has length $\overline{L}$ and bending stiffness $EI$. The $\overline{x}$-axis coincides with the undeformed centerline of the beam, with origin at mid-span. The beam is subjected to axial, compressive end loading $P$, and we denote the small transverse displacement of the beam by $\overline{w}(\overline{x})$. The total potential energy of the system is given by

\begin{equation}\label{eq:TPE}
\overline{\mathcal{E}}[\overline{w}] =\int _{ -\overline{L}/2}^{\overline{L}/2}\left [\frac{1}{2}\genfrac{(}{)}{}{}{d^{2}\overline{w}}{d\overline{x}^{2}}^{2} -\frac{1}{2}P\genfrac{(}{)}{}{}{d\overline{w}}{d\overline{x}}^{2} +\frac{1}{2}k_{2}\overline{w}^{2} -\frac{1}{4}k_{4}\overline{w}^{4}\right ]d\overline{x} .
\end{equation}
\\We rescale via $x =\overline{x}/L_{c}$, $w=\overline{w}/L_{c}$ and $L =\overline{L}/L_{c}$, where $L_{c} : =\left (EI/k_{2}\right )^{1/4}$  is chosen as a characteristic length. The non-dimensional potential energy functional, $\mathcal{E} : =\overline{\mathcal{E}}L_{c}/EI ,$ then reads

\begin{equation}\label{eq:TPE_scaled}
\mathcal{E}[w] =\int _{ -L/2}^{L/2}\left [\frac{1}{2}\left (w^{ \prime  \prime }\right )^{2} -\frac{\lambda }{2}\left (w^{ \prime }\right )^{2} +\frac{1}{2}w^{2} -\frac{1}{4}w^{4}\right ]dx ,
\end{equation}
\\where $( \centerdot )^{ \prime } : =\frac{d( \centerdot )}{dx}$, $\lambda  : =\frac{P}{\sqrt{k_{2}EI}}$ and $k_{4}$$ \equiv k_{2}\sqrt{\frac{k_{2}}{EI}}$.           

\section{Bifurcation Analysis}\label{sec:Section_3}

\subsection{Primary Bifurcation}\label{sec:Section_3_1}

The first variation of the energy \eqref{eq:TPE_scaled} leads to the equilibrium equation

\begin{equation}\label{eq:EL_w(x)}
w^{ \prime  \prime  \prime \prime } +\lambda w^{ \prime  \prime } +w -w^{3} =0 ,
\end{equation}
\\on $\left ( -L/2 ,L/2\right ) ,$ subject to the boundary conditions

\begin{equation}\label{eq:BCs_w(x)}
w( \pm L/2) =w^{ \prime  \prime }( \pm L/2) =0.
\end{equation}
\par We view system \eqref{eq:EL_w(x)} and \eqref{eq:BCs_w(x)} as defining a continuously differentiable mapping $F( \cdot )$ from all real numbers $\lambda $ and all four-times continuously differentiable functions $w$ on $\left [ -L/2 ,L/2\right ]$ that satisfy \eqref{eq:BCs_w(x)} into all continuously differentiable functions on $\left [ -L/2 ,L/2\right ]$ (with the usual maximum norms). Thus, \eqref{eq:EL_w(x)}, \eqref{eq:BCs_w(x)} is equivalent to $F(\lambda, w) =0$. Clearly the straight state $w \equiv 0$ gives a trivial solution, i.e., $F(\lambda, 0) \equiv 0$.

The linearization of  \eqref{eq:EL_w(x)} about the trivial solution is given by

\begin{equation}\label{eq:Linearized_w(x)=0}
T(\lambda )w : =w^{ \prime  \prime  \prime  \prime } +\lambda w^{ \prime  \prime }+w=0,
\end{equation}
\\subject to \eqref{eq:BCs_w(x)}. Here the linear operator $T(\lambda ) =D_{w}F(\lambda, 0)$ is the total (Fr{\'e}chet) derivative of $F(\lambda, \cdot )$ evaluated at $w =0$. We focus here on two possible families of nontrivial solutions of the linearized problem, each at $\lambda=2$: 

Symmetric:

\begin{equation}\label{eq:Symmetric_Length}
L =L_{s} : =(2n -1)\pi ,n =1 ,2 , . . .;\ w=h_{s} : =\sqrt{\frac{2}{L_{s}}}\cos (x);
\end{equation}

Anti-symmetric:

\begin{equation}\label{eq:Antisymmetric_Length}
L =L_{a} : =2n\pi, n =1 ,2 , . . .;\ w = h_{a} : =\sqrt{\frac{2}{L_{a}}}\sin (x) .
\end{equation}

It's not hard to show that $T(\lambda )$ is self-adjoint, and the usual sufficient condition for bifurcation \cite{Crandall_Rabinowitz} at $(\lambda,w) =(2 ,0)$ is satisfied, viz., $\left \langle h_{\beta } ,T^{ \prime }(2)h_{\beta }\right \rangle  \neq 0 ,\beta=s ,a ,$ where $\left \langle f ,g\right \rangle  : =\int _{ -L/2}^{L/2}f(x)g(x)dx$. Accordingly, we deduce the existence of nontrivial solutions of \eqref{eq:EL_w(x)}, \eqref{eq:BCs_w(x)}, and the usual Taylor-series expansion (e.g., \cite{Iooss_Joseph}) reveals sub-critical pitchfork bifurcations:

Symmetric:

\begin{equation}\label{eq:Taylor_Series_Primary_Symm}
\lambda=2 -\frac{3}{2L_{s}}\eta ^{2} +\mathcal{O}(\eta ^{4}) ,\  w_{s} =\eta \sqrt{\frac{2}{L_{s}}}\cos (x) +\mathcal{O}(\eta ^{3});
\ \eta  : =\left \langle h_{s} ,w\right \rangle  \rightarrow 0;
\end{equation}

Anti-symmetric:

\begin{equation}\label{eq:Taylor_Series_Primary_Antisymm}
\lambda  =2 -\frac{3}{2L_{a}}\eta ^{2} +\mathcal{O}(\eta ^{4}) ,\  w_{a} =\eta \sqrt{\frac{2}{L_{a}}}\sin (x) +\mathcal{O}(\eta ^{3});
\ \eta  : =\left \langle h_{a} ,w\right \rangle  \rightarrow 0.
\end{equation}

\subsection{Secondary Bifurcation}\label{sec:Section_3_2}

We now focus on secondary bifurcation from the anti-symmetric solution branch \eqref{eq:Taylor_Series_Primary_Antisymm} in a small half-neighborhood of $\lambda =2$. In particular, we choose $\varepsilon  =\sqrt{2 -\lambda }$ and $L =L_{a}$, and (following \cite{Potier-Ferry_1}) seek a 2-scale solution of the form

\begin{equation}\label{eq:Ansatz_Antisymmetric}
w_{a} =\varepsilon A(X)\sin (x) ,
\end{equation}
\\where $X =\varepsilon x$. A comparison of \eqref{eq:Taylor_Series_Primary_Antisymm} and \eqref{eq:Ansatz_Antisymmetric} implies that 

\begin{equation}\label{eq:w(x)_epsilon_relation_Primary}
\eta  =\sqrt{\frac{2L_{a}}{3}}\varepsilon  \Longrightarrow w_{a} =\varepsilon \frac{2}{\sqrt{3}}\sin (x) +\mathcal{O}(\varepsilon ^{3}).
\end{equation}
\\It will soon be clear that a similar analysis can be carried out for $w_{s}$ with $L =L_{s}$. In any case, \eqref{eq:Ansatz_Antisymmetric} and the boundary conditions \eqref{eq:BCs_w(x)} imply that

\begin{equation}\label{eq:BCs_A(X)}
\frac{dA}{dX} \mid _{X = \pm \frac{\varepsilon L_{a}}{2}} =0.
\end{equation}
\\Next we substitute \eqref{eq:Ansatz_Antisymmetric} into the energy \eqref{eq:TPE_scaled}. To leading order in $\varepsilon$, integration by parts and \eqref{eq:BCs_A(X)} yield 

\begin{equation}\label{eq:TPE3}
\mathcal{E}_{a}[A] =\varepsilon ^{3}\int _{ -\frac{\varepsilon L_{a}}{2}}^{\frac{\varepsilon L_{a}}{2}}\left [\genfrac{(}{)}{}{}{dA}{dX}^{2} +\frac{1}{4}A^{2} -\frac{3}{32}A^{4}\right ]dX ,
\end{equation}  
\\the first variation of which delivers the amplitude equation

\begin{equation}\label{eq:Amplitude_Equation}
\frac{d^{2}A}{dX^{2}} -\frac{1}{4}A +\frac{3}{16}A^{3} =0.
\end{equation}
\\In this case the trivial solution of \eqref{eq:Amplitude_Equation} subject to \eqref{eq:BCs_A(X)} is a constant, say, $A_{o}$. Comparing \eqref{eq:Ansatz_Antisymmetric} and \eqref{eq:w(x)_epsilon_relation_Primary}, we see that $A_{o} =\frac{2}{\sqrt{3}}$. We then look for a nontrivial solution of the form $A(\varepsilon x) =\frac{2}{\sqrt{3}}+ u(x)$, which leads to the boundary value problem

\begin{equation}\label{eq:EL_u(x)}
u^{ \prime  \prime } +\frac{(2 -\lambda )}{2}\left (u +\frac{3\sqrt{3}}{4}u^{2} +\frac{3}{8}u^{3}\right ) =0 ,u^{ \prime }( \pm L_{a}/2) =0.
\end{equation}
\\Clearly $u \equiv 0$ is the trivial solution of \eqref{eq:EL_u(x)}, and the linearized problem is then given by

\begin{equation}\label{eq:Linearized_u(x)=0}
u^{ \prime  \prime } +\frac{(2 -\lambda )}{2}u =0 ,
\end{equation}
\\subject to the same boundary conditions \eqref{eq:BCs_A(X)}. This problem admits the nontrivial solution $u =C\cos \left (\gamma _{k}x\right )$ at $\lambda=2 -2\left (\gamma _{k,a}\right )^{2}$, where $\gamma _{k,a} : =\frac{2k\pi }{L_{a}}, k =1 ,2 , \dots$. Then as before in the previous section, a Taylor-series expansion yields the following family of sub-critical pitchfork bifurcating solution branches for \eqref{eq:EL_u(x)}:

\begin{equation}\label{eq:Taylor_Series_Secondary}
\lambda  =2 -2\left (\gamma _{k,a}\right )^{2} -\frac{9}{2L_{a}}\left (\gamma _{k,a}\right )^{2}\eta ^{2} +\mathcal{O}\left (\eta ^{3}\right ),u =\eta \sqrt{\frac{2}{L_{a}}}\ensuremath{\operatorname*{}}\cos \left (\gamma _{k,a}x\right ) +\mathcal{O}\left (\eta ^{2}\right ) ,
\end{equation}
\\as $\eta  \rightarrow 0, k =1 ,2 , \dots$. Since $A(X) =\frac{2}{\sqrt{3}} + u(X/\varepsilon )$, we see that \eqref{eq:Taylor_Series_Secondary} represents a family of solutions of \eqref{eq:BCs_A(X)}, \eqref{eq:Amplitude_Equation} bifurcating from the constant solution $A =\frac{2}{\sqrt{3}}$.

\section{Asymptotic Analysis}\label{sec:Asymp_Analysis}\label{sec:Section_4}

In this section we return to the amplitude equation \eqref{eq:Amplitude_Equation} the general solution of which can be expressed in terms of the Jacobi elliptic ``dn'' function, viz.,

\begin{equation}\label{eq:Jacobi_dn}
A(X) =\alpha \ \mathrm{dn}(\Omega X ,m) ,m \in [0 ,1) .
\end{equation}
\\Our goal here is to obtain an asymptotic approximation of $A$ as the length of the beam becomes very large but stays finite. Substituting \eqref{eq:Jacobi_dn} into \eqref{eq:Amplitude_Equation}, we find

\begin{equation}\label{eq:alpha(m)_Omega(m)}
\alpha=2\sqrt{\frac{2}{3(2 -m)}} ,\Omega  =\frac{1}{2\sqrt{2 -m}}.
\end{equation}

\begin{figure}[H]
    \centering
        \includegraphics[width=0.85\textwidth]{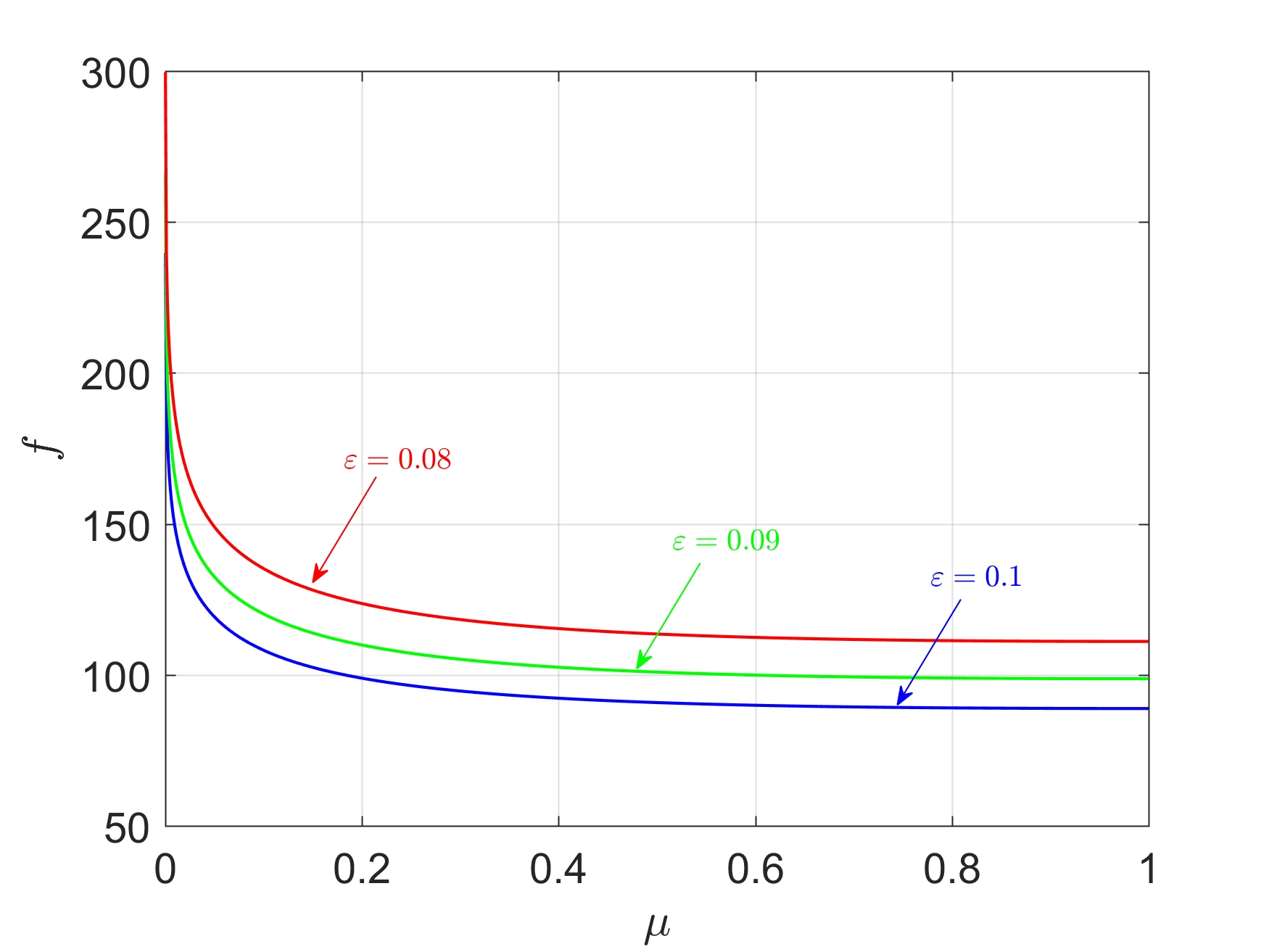}
   \caption{}
   \label{fig:f(mu)}
\end{figure}

Next we use the boundary conditions \eqref{eq:BCs_A(X)} to deduce

\begin{equation}\label{eq:BCs_sn_cn}
\alpha \ m \ \Omega \ \mathrm{sn}\genfrac{(}{)}{}{}{\varepsilon \Omega L_{a}}{2} \ \mathrm{cn}\genfrac{(}{)}{}{}{\varepsilon \Omega L_{a}}{2} =0.
\end{equation}
\\At $m =0$, $\mathrm{dn}(\Omega X ,0) \equiv 1$, and from \eqref{eq:Amplitude_Equation}, \eqref{eq:alpha(m)_Omega(m)}, the amplitude reduces to $A \equiv 2/\sqrt{3}$, which we recall represents the primary bifurcating path, cf. after \eqref{eq:Taylor_Series_Secondary}. Thus, we require $m \in (0 ,1)$ for secondary bifurcation. With this in hand $(m \neq 0)$, \eqref{eq:BCs_sn_cn} implies that the argument of $\mathrm{sn}$ or $\mathrm{cn}$ is equal to a quarter period, viz.,  

\begin{equation}\label{eq:Quarter_period}
\varepsilon \Omega L_{a} =2K(m) ,
\end{equation}
\\where 

\begin{equation}\label{eq:Elliptic_integral}
K(m) =\int _{0}^{\pi /2}\frac{1}{\sqrt{1 -m\sin ^{2}\phi }}d\phi,
\end{equation}
\\cf. \cite{Abramowitz_Stegun}. We note here that $K(m) \longrightarrow \infty $ as $m \nearrow 1$.
\\We now fix the $\varepsilon>0$ sufficiently small, and seek an expression for $L_{a}$. From  \eqref{eq:alpha(m)_Omega(m)} and \eqref{eq:Quarter_period} we find

\begin{equation}\label{eq:f_function}
L_{a} =f(\mu ) : =\frac{4\sqrt{1 +\mu }}{\varepsilon }K(1 -\mu ),
\end{equation}
\\where $\mu  : =1-m, \mu \in (0 ,1)$, is the complementary modulus. From \eqref{eq:f_function} and the behavior of $K$, we see that

\begin{equation}\label{eq:f_mu_limits}
f \longrightarrow \infty \ as\ \mu  \searrow 0 ,\ and\ f(1) =\frac{2\sqrt{2}\pi }{\varepsilon } .
\end{equation}
\\Moreover, we deduce

\begin{equation}\label{eq:f_diff_mu}
\frac{df}{d\mu } =g(\mu ) ,
\end{equation}
\\where

\begin{equation}\label{eq:g(mu)}
g(\mu ) = -\frac{2}{\varepsilon \mu \sqrt{1 +\mu }}\int _{0}^{\pi /2}\frac{\sin ^{2}(\phi ) +\mu \cos ^{2}(\phi )}{\sqrt{\cos ^{2}(\phi ) +\mu \sin ^{2}(\phi )}}d\phi<0,
\end{equation}
\\i.e., $f$ is monotonically decreasing. The above observations can be verified by plotting $f$ as a function of $\mu$ for fixed values of $\epsilon$ as shown in Fig. \ref{fig:f(mu)}.

Hence, \eqref{eq:f_mu_limits}-\eqref{eq:g(mu)} imply there is a unique value $\mu  \in (0 ,1)$ for every 

\begin{equation}\label{eq:L_cr}
L_{a}>\frac{2\sqrt{2}\pi }{\varepsilon },
\end{equation}
\\i.e., secondary solutions exist. We note that at criticality (equality), \eqref{eq:L_cr} gives $\epsilon_{c}=2\sqrt{2}\pi/L_a$, which agrees with the bifurcation value from \eqref{eq:Linearized_u(x)=0}, viz., $\lambda_{c}=2-\epsilon_{c}^{2}=2-8\pi^{2}/L_{a}^{2}$, cf. \eqref{eq:Linearized_u(x)=0}, \eqref{eq:Taylor_Series_Secondary}.

Of course \eqref{eq:f_function}-\eqref{eq:g(mu)} also imply that $\mu  =f^{ -1}(L_{a})$, where $f^{ -1}$ is monotonically decreasing for  $L_{a} \in (\frac{2\pi \sqrt{2}}{\varepsilon } ,\infty )$, with $f^{ -1}(\genfrac{(}{)}{}{}{2\pi \sqrt{2}}{\varepsilon })=1$ and $f^{ -1} \searrow 0\overset{}{}$  as $L_{a}$$ \longrightarrow \infty$. Accordingly, we seek an asymptotic solution of \eqref{eq:f_function} in the limit as $\mu $ goes to zero. From \cite{Abramowitz_Stegun} we observe that $K$ (cf. \eqref{eq:Elliptic_integral}) has a logarithmic singularity at $\mu =0$, from which we deduce

\begin{equation}\label{eq:Asymptotic_mu}
L_{a} =f(\mu ) \sim \frac{2}{\varepsilon }\ln (\genfrac{(}{)}{}{}{16}{\mu })\ \text{as} \ \mu  \searrow 0 \Longrightarrow \mu  \sim 16 \ \exp\left( -\frac{\varepsilon L_{a}}{2}\right)\ \text{as} \ L_{a} \longrightarrow \infty. 
\end{equation}

We now obtain an asymptotic expression for $A(X)$ from the \eqref{eq:Jacobi_dn} when $\mu$ is close to zero. Define $A^*(X): =\lim_{\mu \searrow 0} \left[\alpha\ \mathrm{dn}(\Omega X,\mu)\right]$. Noting that $\lim_{\mu\searrow0}\alpha(\mu)=2\sqrt{2/3}$ and $\lim_{\mu\searrow0}\Omega(\mu)=1/2$, one obtains,

\begin{equation}\label{eq:Asymptotic_ASymm_Amplitude_limit}
\begin{split}
A^*(X)&=2\sqrt{\frac{2}{3}} \mathrm{dn}\left(\frac{X}{2},\mu\searrow0\right).
\end{split}
\end{equation}
\\The $\mathrm{dn(X/2,\mu)}$ function can be approximated in terms of hyperbolic functions when $\mu$ is close to zero \cite{Abramowitz_Stegun}, in which case $A^*(X)$ has an asymptotic expression given by,

\begin{equation}\label{eq:Asymptotic_ASymm_Amplitude_hyperbolic}
\begin{split}
A^*(X)\sim2\sqrt{\frac{2}{3}}\ \mathrm{sech}\left(\frac{X}{2}\right) \left[1+\frac{\mu}{4}\left\{\sinh\left(\frac{X}{2}\right) \cosh\left(\frac{X}{2}\right)+\left(\frac{X}{2}\right)\right\}\tanh\left(\frac{X}{2}\right)\right].
\end{split}
\end{equation}
\\Finally, on substituting $\mu$ from \eqref{eq:Asymptotic_mu} in \eqref{eq:Asymptotic_ASymm_Amplitude_hyperbolic} we get,

\begin{equation}\label{eq:Asymptotic_ASymm_Amplitude}
\begin{split}
A^*(X)&\sim2\sqrt{\frac{2}{3}}\ \mathrm{sech}\left(\frac{X}{2}\right) \\ &\left[1+4\exp\left({-\varepsilon L_a}/{2}\right)\left\{\sinh\left(\frac{X}{2}\right) \cosh\left(\frac{X}{2}\right)+\left(\frac{X}{2}\right)\right\}\tanh\left(\frac{X}{2}\right)\right].
\end{split}
\end{equation}

If we begin as in \eqref{eq:Ansatz_Antisymmetric}, but now with the symmetric ansatz $w_{s}=\varepsilon A(X) \cos(x)$ for $L=L_{s}$ (cf. \eqref{eq:Taylor_Series_Primary_Symm}), it's not hard to see that the amplitude equation \eqref{eq:Amplitude_Equation} is unchanged on the domain $(-L_{s}/2,L_{s}/2)$. Likewise, the boundary conditions \eqref{eq:BCs_A(X)} hold at $X=\pm\varepsilon L_{s}/2$. Carrying out precisely the same steps as in Section \ref{sec:Section_3_2}, we arrive at a family of secondary pitchfork bifurcating solutions, sub-critically bifurcating from the primary solution \eqref{eq:Taylor_Series_Primary_Symm}, of the same form as \eqref{eq:Taylor_Series_Secondary}:
\begin{equation}
\lambda  =2 -2\left (\gamma _{k,s}\right )^{2} -\frac{9}{2L_{s}}\left (\gamma _{k,s}\right )^{2}\eta ^{2} +\mathcal{O}\left (\eta ^{3}\right),
u =\eta \sqrt{\frac{2}{L_{s}}}\ensuremath{\operatorname*{}}\cos \left (\gamma _{k,s}x\right ) +\mathcal{O}\left (\eta ^{2}\right ) ,
\end{equation}
\\where $\gamma_{k,s}:=\frac{2k\pi}{L_s}, k=1,2,\dots$. Moreover, an analysis identical to that given here in Section \ref{sec:Asymp_Analysis}, gives the same asymptotic expression \eqref{eq:Asymptotic_ASymm_Amplitude} for the amplitude, but with $L_s$ in place of $L_a$. Of course, these two amplitude functions have the same limit as $L_{s},L_{a}\to\infty$, viz.,

\begin{equation}
A^{*}(X)\to2\sqrt{\frac{2}{3}} \sech\left(\frac{X}{2}\right),
\end{equation}
\\which confirms the fact that an infinitely long beam admits a 1-parameter family of solutions
\begin{equation}
w_{\infty,\phi}=2\sqrt{\frac{2}{3}} \sech\left(\frac{X}{2}\right)\sin\left(x-\phi\right), \phi\in[0,2\pi).
\end{equation}
\\Our goal is to show that this is essentially the case for extremely long beams as well.

We start with a beam of length $L_{a}=2n\pi$, for very large values of $n$, and consider the asymptotic solution $w_{a}(x)=\varepsilon A^{*}(X)\sin(x)$, with $\varepsilon=\sqrt{2-\lambda}$ sufficiently small and fixed. In the view of \eqref{eq:BCs_A(X)}, $w_{a}$ satisfies the boundary conditions \eqref{eq:BCs_w(x)}. Now define
\begin{equation}\label{eq:w_a_phi}
w_{a,\phi}=\varepsilon A^{*}(X)\sin\left(x-\phi\right), \phi\in[0,2\pi),
\end{equation}
\\which is also an asymptotic solution that does not satisfy the boundary conditions, unless $\phi=0$. However, for sufficiently large values of $L_a$ (or $n$), \eqref{eq:Asymptotic_ASymm_Amplitude} yields
\begin{equation}\label{eq:Difference_Amplitude}
\begin{split}
&w_{a,\phi} \left(x=\frac{L_a}{2}\right)\sim2\varepsilon\sqrt{\frac{2}{3}} \ \mathrm{sech}\left(\varepsilon\frac{L_a}{4}\right)\sin\left(\frac{L_a}{2}-\phi\right) \\ &\left[1+4\exp\left({-\varepsilon L_a}/{2}\right)\left\{\sinh\left(\varepsilon\frac{L_a}{4}\right) \cosh\left(\varepsilon\frac{L_a}{4}\right)+\left(\varepsilon\frac{L_a}{4}\right)\right\}\tanh\left(\varepsilon\frac{L_a}{4}\right)\right] \\ &=2\varepsilon\sqrt{\frac{2}{3}} \ \mathrm{sech}\left(\varepsilon\frac{L_a}{4}\right)\sin\left(\frac{L_a}{2}-\phi\right) \\ & \left[2+\exp\left({-\varepsilon L_a}\right)+\exp\left({-\varepsilon L_a}/{2}\right)\left(\varepsilon L_a \tanh\left(\varepsilon\frac{L_a}{4}\right)-2\right)\right].
\end{split}
\end{equation}
\\with a similar expression resulting at $x=-L_a/2$. That is, for $\phi\ne0$, the end displacements miss satisfying the boundary conditions by exponentially small terms only. Clearly, we can start with $w_{s}(x)=\varepsilon A^{*}(X)\cos(x)$, and make the same argument for $w_{s,\phi}(x)=\varepsilon A^{*}(X)\cos(x-\phi)$, for an arbitrary phase shift $\phi$, and arrive at the same conclusion for very large $L_{s}$.

In a similar manner, we substitute \eqref{eq:Jacobi_dn} into \eqref{eq:TPE3} to obtain an asymptotic expression for the potential energy:

\begin{equation}
\begin{split}
\mathcal{E}_a=\frac{2\varepsilon^3}{3}\int_{-\frac{\varepsilon L_a}{2}}^{\frac{\varepsilon L_a}{2}}&\Bigg[\frac{m^2}{(2-m)^2} \mathrm{sn}^2(\Omega X,m)\ \mathrm{cn}^2(\Omega X,m) \\  &+\frac{1}{(2-m)}\mathrm{dn}^2(\Omega X,m)-\frac{1}{(2-m)^2}\mathrm{dn}^4(\Omega X,m)\Bigg]dX.
\end{split}
\end{equation}
\\ Since the integrand of $\mathcal{E}_a$ is even, the energy can be rewritten via the change of variable $\Omega X=t$, which yields

\begin{equation}\label{eq:TPE_var_t}
\begin{split}
\mathcal{E}_a=\frac{8\varepsilon^3}{3}\int_{0}^{\frac{\varepsilon \Omega L_a}{2}}&\Bigg[\frac{m^2}{(2-m)^\frac{3}{2}} \mathrm{sn}^2(t)\ \mathrm{cn}^2(t) +\frac{1}{(2-m)^\frac{1}{2}}\mathrm{dn}^2(t)\\  &-\frac{1}{(2-m)^\frac{3}{2}}\mathrm{dn}^4(t)\Bigg]dt.
\end{split}
\end{equation}
\\We now substitute $m=1-\mu$ in the expression of the total potential energy $\mathcal{E}_a$ given into \eqref{eq:TPE_var_t}, and then expand the integrand in a Taylor series centered around $\mu=0$ . On approximating the functions $\mathrm{sn(t)}$, $\mathrm{cn(t)}$ and $\mathrm{dn(t)}$ in terms of hyperbolic functions for $\mu \searrow 0$ \cite{Abramowitz_Stegun} and neglecting the terms of order $\mu^2$ and higher, we find 

\begin{align}\label{eq:Asymptotic_Energy}
\mathcal{E}_a^*\sim\mathcal{E}_\infty+\mathcal{E}_{L_a},
\end{align}
where,
\begin{align}\label{eq:Asymptotic_Energy_infty}
\mathcal{E}_\infty &=\varepsilon^3\left(\frac{8}{3}\right)\int_{0}^{\frac{\varepsilon \Omega L_a}{2}}\frac{2 \ \mathrm{sinh}^2(t)}{\mathrm{cosh}^4(t)}dt =\varepsilon^3\left(\frac{16}{9}\right),
\end{align}
\\and
\begin{equation}\label{eq:Asymptotic_Energy_L_a}
\begin{split}
\mathcal{E}_{L_a} &=\varepsilon^3 \ \mu \left(\frac{8}{3}\right) \int_{0}^{\frac{\varepsilon \Omega L_a}{2}}\left[\frac{t\ \mathrm{sinh}(t)}{\mathrm{cosh}^3(t)}-\frac{2 \ t \ \mathrm{sinh}(t)}{\mathrm{cosh}^5(t)}+\frac{5}{\mathrm{cosh}^4(t)}-\frac{4}{\mathrm{cosh}^2(t)}\right]dt \\ &=- \varepsilon^3\exp{\left(-{\varepsilon L_a}/{2}\right)}\left(\frac{4}{3}\right).
\end{split}
\end{equation}
\\ Here $\mathcal{E}_\infty$ corresponds to the energy of an infinitely large beam and $\mathcal{E}_{L_a}$ represents the energy correction owing to the finiteness of the beam length. For a symmetric ansatz $w_{s}=\varepsilon A(X) \cos(x)$ with $L=L_s$, it can be shown that the asymptotic expression for energy is identical to \eqref{eq:Asymptotic_Energy}, with $L=L_{s}$. When $L_{a}, L_{s}\to\infty$, $\mathcal{E}_{L_a}=\mathcal{E}_{L_s}=0$, and we recover the energy of an infinitely long beam. Moreover, the energy of the asymptotic solution $w_{a,\phi}$ can be obtained by substituting \eqref{eq:w_a_phi} in \eqref{eq:TPE_scaled}. Noting that $A^*(X)$ satisfies \eqref{eq:BCs_A(X)} at the boundaries $x=\pm L_{a}/2$, an integration of \eqref{eq:TPE_scaled} by parts gives the leading order terms (go to \eqref{eq:Anti-symm_Energy_phi})

\begin{equation}\label{eq:Anti-symm_Energy_phi}
\mathcal{E}_{a,\phi}\sim-\varepsilon ^{2}\left[\frac{\left(A^*\left(\varepsilon x\right)\right)^2 \sin\left(2\left(x-\phi\right)\right)}{2}\right]\Bigg|_{-\frac{L_{a}}{2}}^{\frac{L_{a}}{2}}+\mathcal{E}_{a},
\end{equation}  
\\where $\mathcal{E}_{a}$ is the energy of the anti-symmetric configuration, cf. \eqref{eq:TPE3}. We observe that the boundary terms in \eqref{eq:Anti-symm_Energy_phi} are exponentially small for fixed, small $\varepsilon>0$ and $L_a$ sufficiently large, cf. \eqref{eq:TPE_var_t}. That is, the energies of beam configurations \eqref{eq:w_a_phi} differ by exponentially small terms for very long length beams. \\
\\
{\bf{Remark}}: As special cases of \eqref{eq:Anti-symm_Energy_phi}, the energies of the anti-symmetric ($\phi=0, \pi$) and the symmetric ($\phi=\pi/2, 3\pi/2$) configurations are recovered. Observe that each of these differ from the energy of an infinitely-long beam by terms exponentially small in beam length, as was conjectured in\cite{Diamant_Witten_2}.
 
\section{Comparison with Numerical Analysis}\label{sec:Section_5}

In this section we present numerical bifurcation results, employing a finite-element model for the beam, with the goal of validating the asymptotic results from the previous section. For the numerical computation, we first consider beam lengths of $40\pi$ and $50\pi$ in the anti-symmetric case, and lengths of $45\pi$ and $55\pi$ in the symmetric case, which are in consonance with \eqref{eq:Symmetric_Length}, \eqref{eq:Antisymmetric_Length}. Later in the section we consider lengths an order of magnitude larger. In choosing these, we note that the asymptotic results of Section \ref{sec:Asymp_Analysis} give strong evidence for the existence of modulated, two-scale solutions for sufficiently small values of the parameter $\varepsilon=\sqrt{2-\lambda}$. We use cubic Hermite shape functions to approximate the displacement of the beam at the element level, and we employ 10 elements for every $\pi$ units of length. Furthermore, we exploit anti-symmetry and symmetry in the numerical calculations. In particular, we consider the system on the interval $(0, L/2)$, with a simple support at $x=L/2$, with the following essential boundary conditions at $x=0$:

\begin{equation}
\begin{split}
&\text{Anti-symmetric case}: w\left(0\right)=0; \\
&\text{Symmetric case}: w'\left(0\right)=0.
\end{split}
\end{equation}

\begin{figure}[H]
    \centering
        \includegraphics[width=0.85\textwidth]{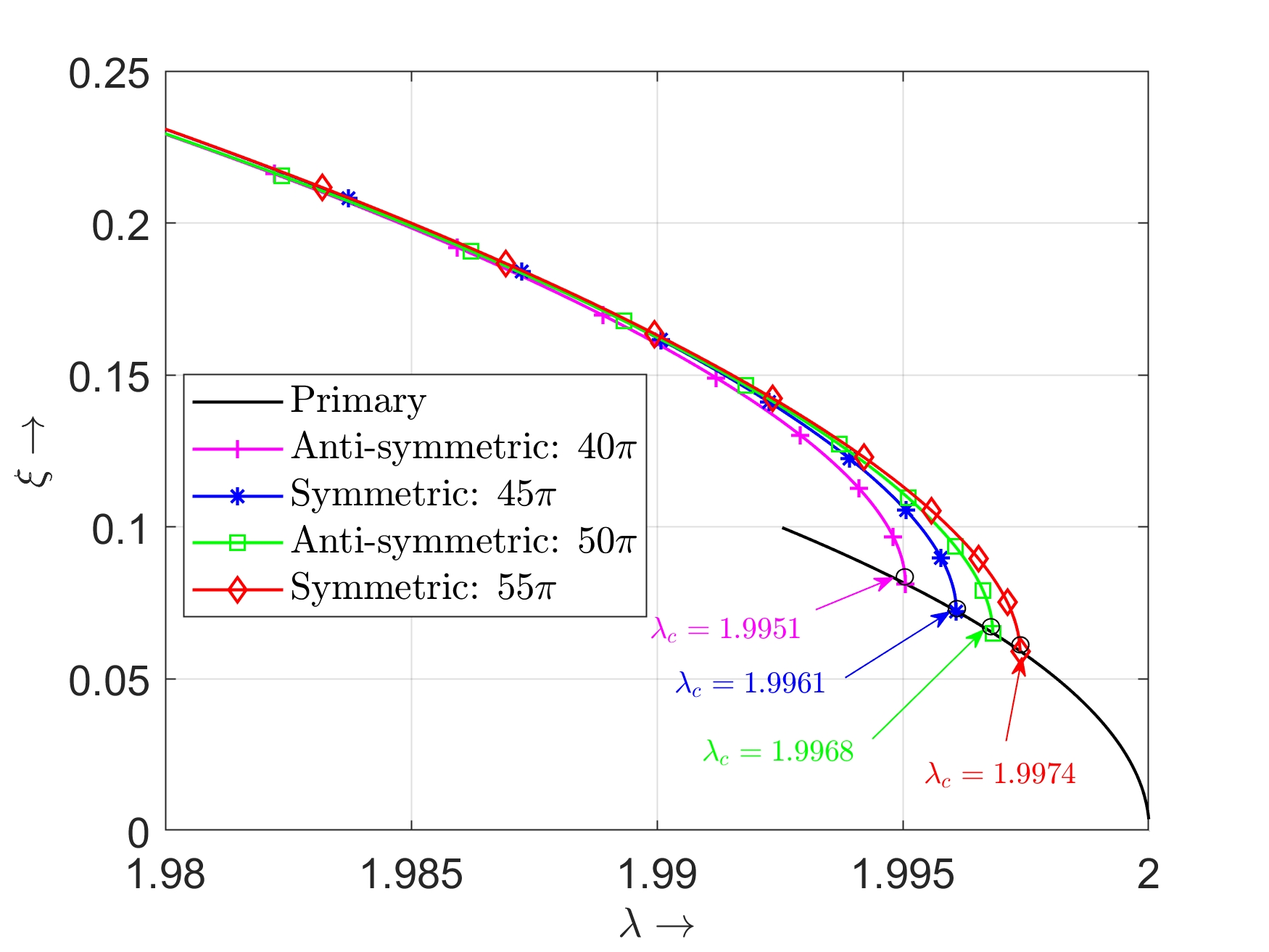}
   \caption{$\xi=\text{max}|w|$ vs. $\lambda=$ load}
   \label{fig:Bifurcation_Diagram}
\end{figure}

We use pseudo-arc-length continuation \cite{Keller} to compute numerical solution paths. We first obtain the primary bifurcation path, corresponding to a sub-critical bifurcation, as depicted in Figure \ref{fig:Bifurcation_Diagram}, where $\xi=\text{max}|w|$ denotes the maximum displacement. All solutions along that path are (extendable to) periodic solutions on the entire $x$-axis. For that reason the solution paths for each of our chosen lengths plot the same in Figure \ref{fig:Bifurcation_Diagram}. Next we compute secondary bifurcating solution paths. We pinpoint the locations of secondary bifurcation points by monitoring the occurence of a zero eigenvalue of the tangent stiffness matrix, and then employ a standard branch-switching technique \cite{Keller} to get onto the secondary bifurcating solution path.

The secondary bifurcation points corresponding to the four lengths are shown as open circles in Figure \ref{fig:Bifurcation_Diagram}. Their computed values are summarized in the Table \ref{table:Secondary_Bifurcation_points} along with their respective values as predicted by the asymptotic analysis, viz., $\lambda_c=2-8\pi^2{L_\alpha}^2 (\alpha=\text{a or s})$, cf. \eqref{eq:L_cr} (and the discussion that immediately follows). The agreement is clearly excellent.

\begin{table}[H]
\begin{center}
\begin{tabular}{ |c|c|c| } 
 \hline
 Beam Length & Analytical & Numerical \\ 
 \hline
 $40\pi$ & $1.9950$ & $1.9951$ \\ 
 \hline
 $45\pi$ & $1.9960$ & $1.9961$ \\ 
 \hline
 $50\pi$ & $1.9968$ & $1.9968$ \\ 
 \hline
 $55\pi$ & $1.9974$ & $1.9974$ \\ 
 \hline
\end{tabular}
\end{center}
 \caption{Secondary Critical loads}
 \label{table:Secondary_Bifurcation_points}
\end{table}

In Table \ref{table:Energy_per_unit_length} we list the computed values of the total potential energy for the lengths considered along with their respective asymptotic values from \eqref{eq:Asymptotic_Energy}-\eqref{eq:Asymptotic_Energy_L_a}. Aside from the very good agreement, we observe that the difference between the asymptotic energy and the numerically calculated energy decreases with an increase in beam lengths.

\begin{table}[H]
\begin{center}
\begin{tabular}{ |c|c|c|c| } 
 \hline
 Beam Length & Analytical $(10^{-5})$ & Numerical $(10^{-5})$ & $\%$ difference \\ 
 \hline
 $40\pi$ & $1.4091$ & $1.3719$ & $2.71$ \\ 
 \hline
 $45\pi$ & $1.2658$ & $1.2473$ & $1.49$ \\ 
 \hline
 $50\pi$ & $1.1296$ & $1.1186$ & $0.98$ \\ 
 \hline
 $55\pi$ & $1.0357$ & $1.0280$ & $0.75$ \\ 
 \hline
\end{tabular}
\end{center}
 \caption{Energy per unit beam length}
 \label{table:Energy_per_unit_length}
\end{table}

\begin{figure}[H]
    \centering
    \begin{subfigure}[H]{0.80\textwidth}
        \centering
        \includegraphics[width=0.90\textwidth]{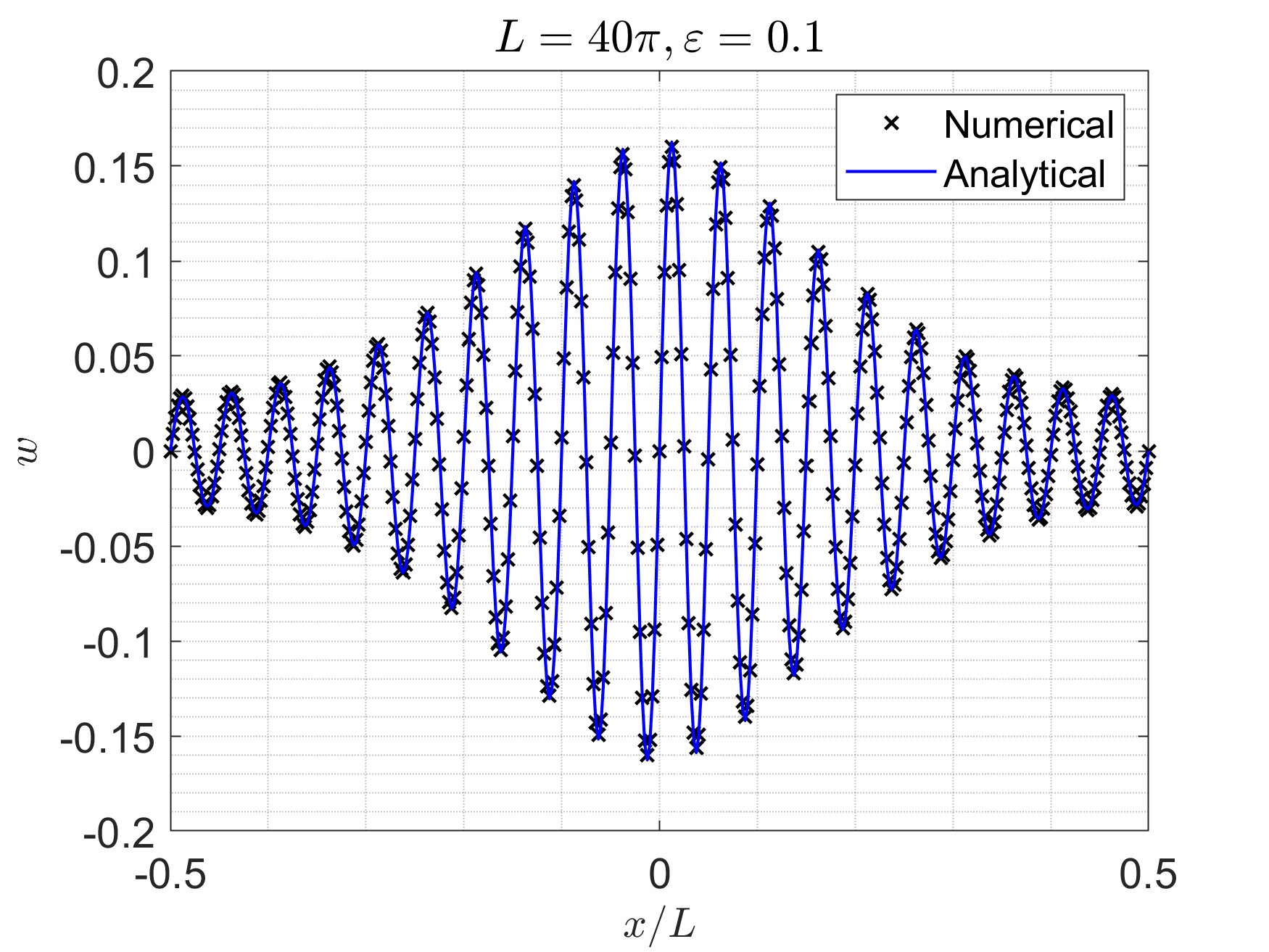}
        \caption{}    
    \end{subfigure}
    \hfill
    \begin{subfigure}[H]{0.80\textwidth}
        \centering
        \includegraphics[width=0.90\textwidth]{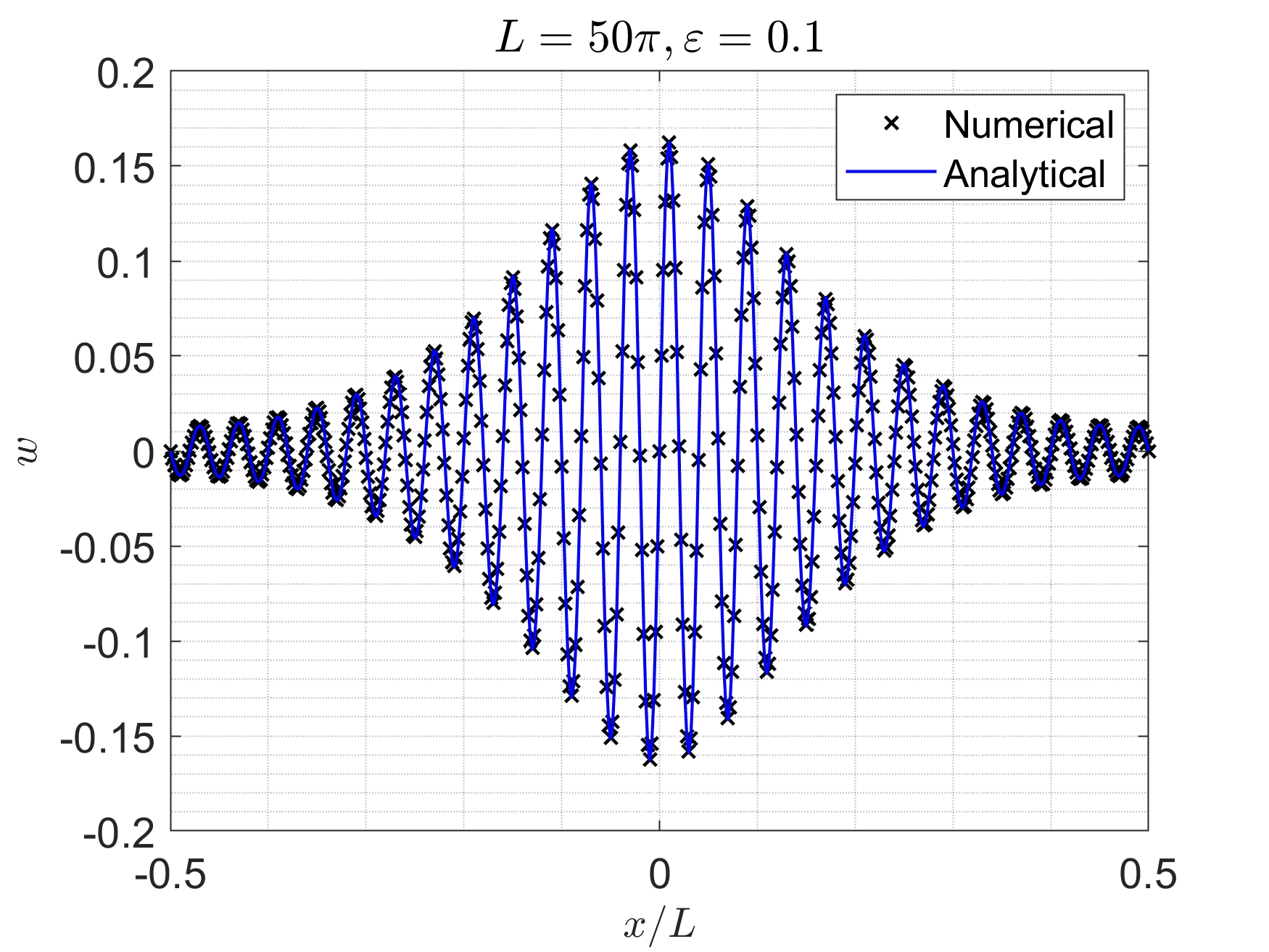}
        \caption{}   
    \end{subfigure}
    \quad
\caption{Comparison for Anti-symmetric Deformation modes}
\label{fig:Antisymm}
\end{figure}

\begin{figure}[H]
    \centering
    \begin{subfigure}[H]{0.80\textwidth}
        \centering
        \includegraphics[width=0.90\textwidth]{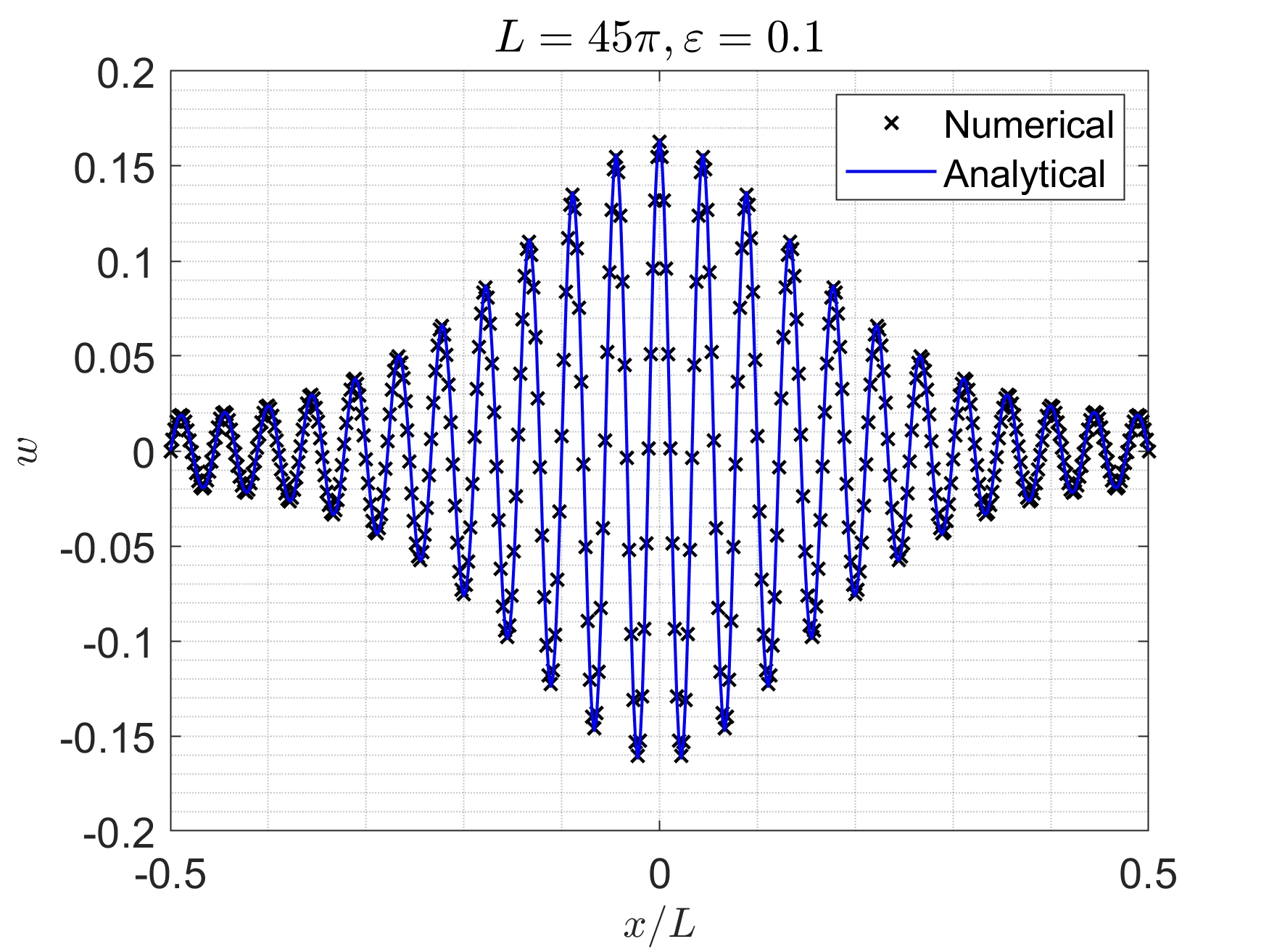}
        \caption{}   
    \end{subfigure}
    \hfill
    \begin{subfigure}[H]{0.80\textwidth}
        \centering
        \includegraphics[width=0.90\textwidth]{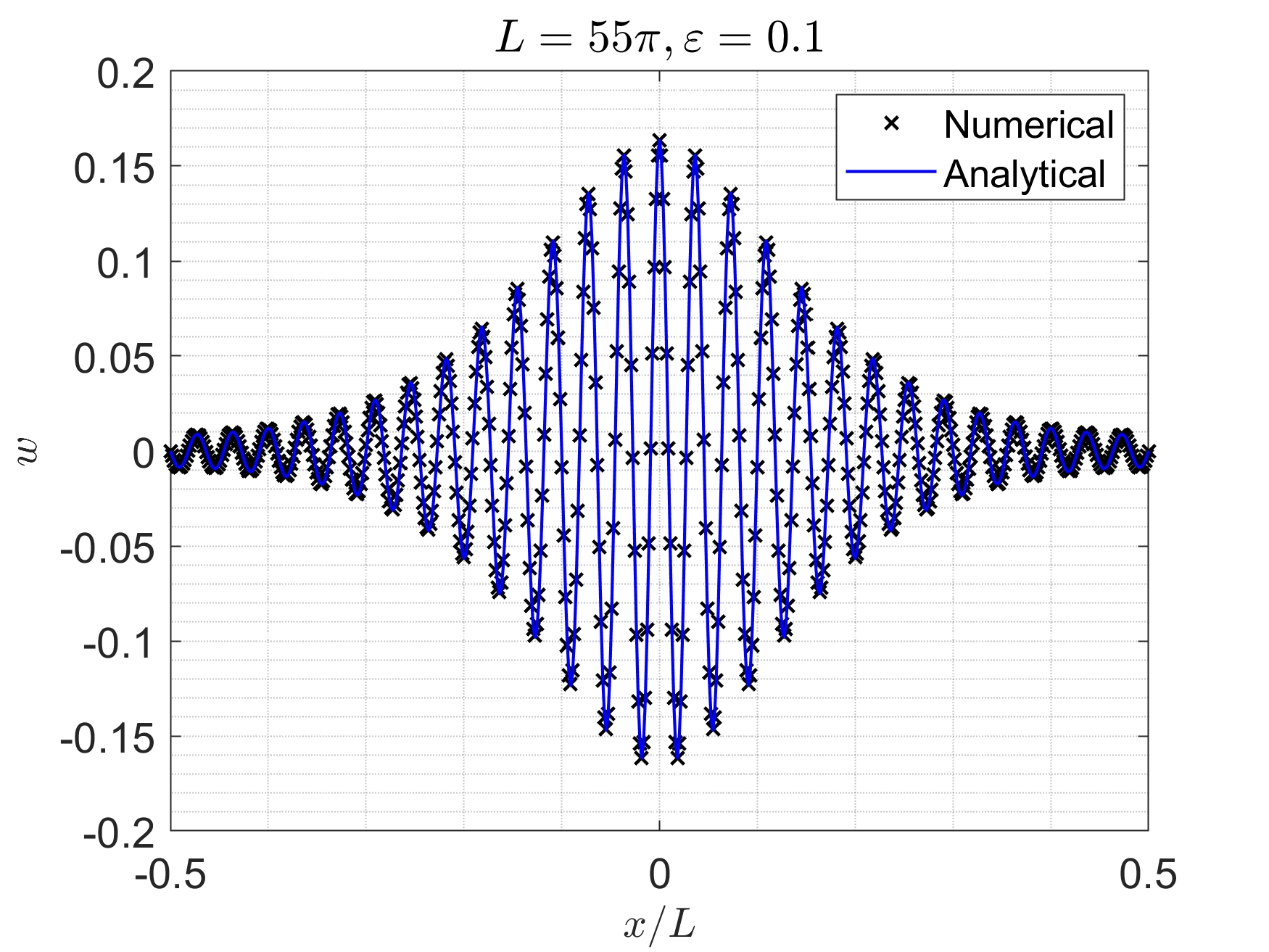}
        \caption{}
    \end{subfigure}
    \quad
\caption{Comparison for Symmetric Deformation modes}
\label{fig:Symm}
\end{figure}

Figures \ref{fig:Antisymm} and \ref{fig:Symm} provide a comparison of the beam deformation calculated using finite element method and the asymptotic analysis, viz., $w(x)=\varepsilon A^*(X)\ p(x)$, $\left(p(x)=\sin(x) \ \text{or}\ \cos(x)\right)$ for anti-symmetric and symmetric modes, respectively, at $\varepsilon=0.1$, which corresponds to $\lambda=1.99$ in Figure \ref{fig:Bifurcation_Diagram}. Once again, the agreement between the numerical results and the asymptotic solutions is excellent.

\begin{figure}[H]
    \centering
        \includegraphics[width=0.80\textwidth]{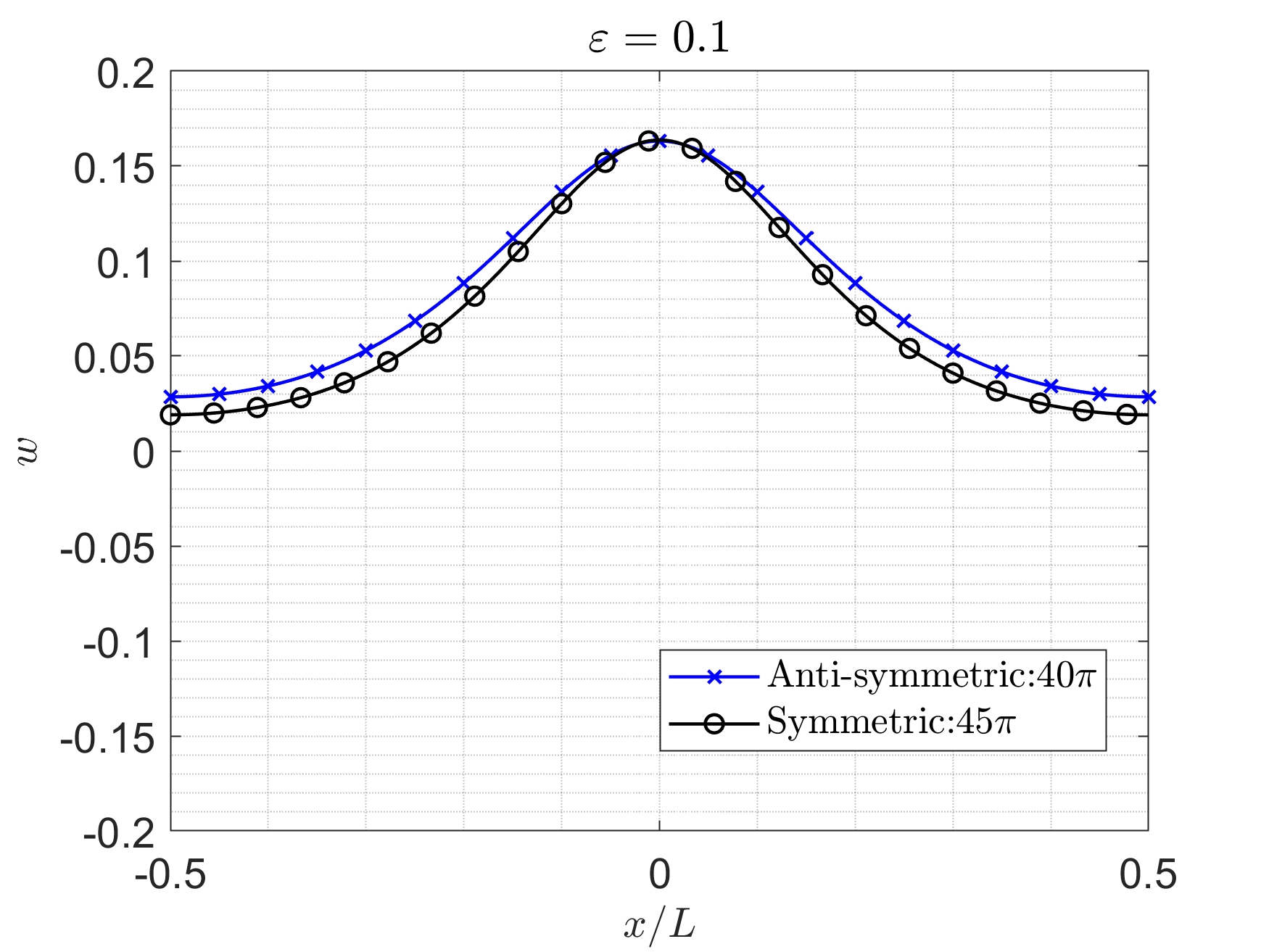}
   \caption{Amplitude Envelope for $L_a=40\pi$ and $L_s=45\pi$}
   \label{fig:Envelope_40_45_eps_0_1}
\end{figure}

\begin{figure}[H]
    \centering
        \includegraphics[width=0.80\textwidth]{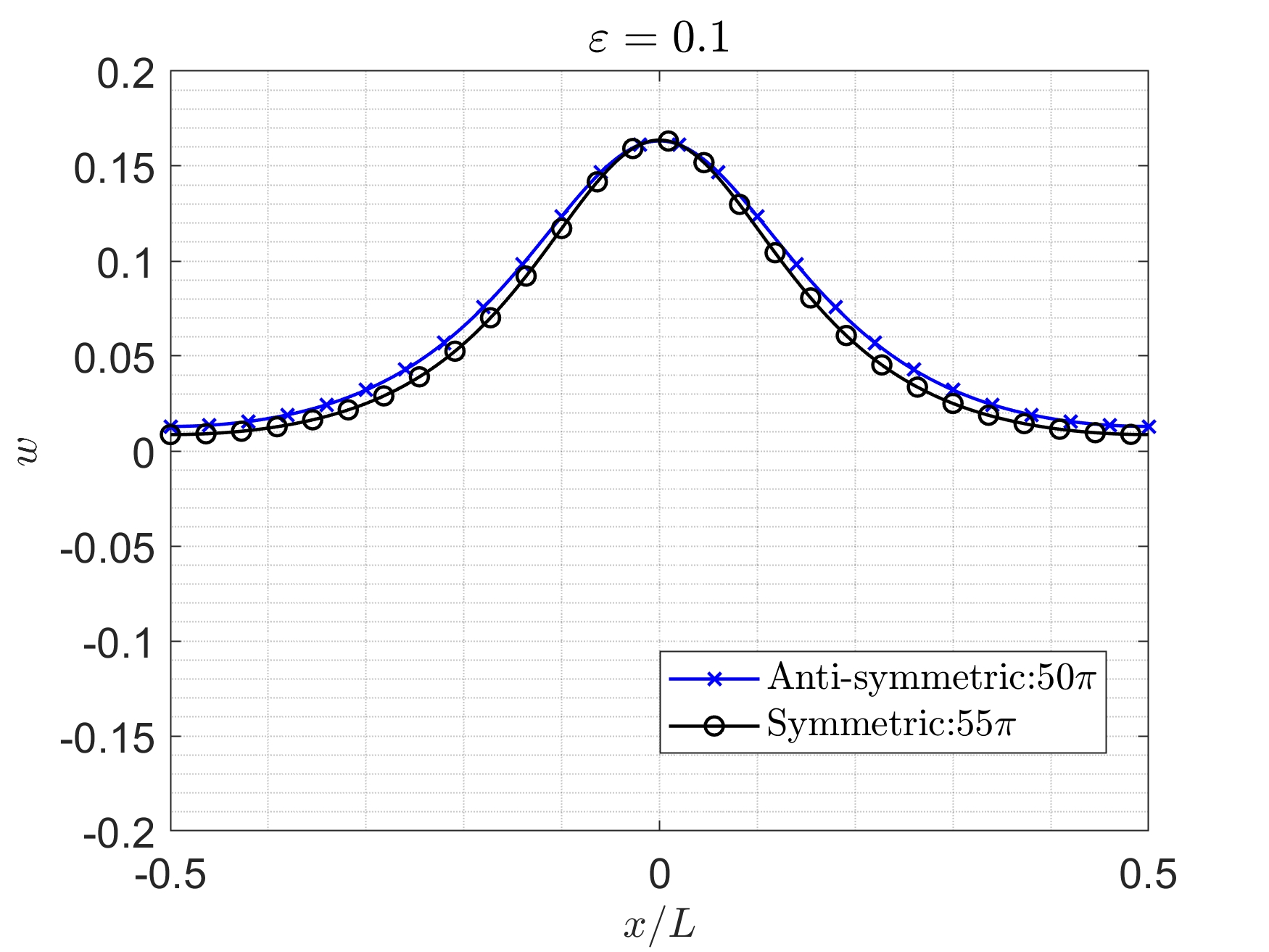}
   \caption{Amplitude Envelope for $L_a=50\pi$ and $L_s=55\pi$}
   \label{fig:Envelope_50_55_eps_0_1}
\end{figure}

Finally, a comparison of the amplitude envelopes obtained from the analytical (asymptotic) expression \eqref{eq:Asymptotic_ASymm_Amplitude} with $L=L_a$ for the anti-symmetric and $L=L_s$ for the symmetric modes is shown in Figures \ref{fig:Envelope_40_45_eps_0_1} and \ref{fig:Envelope_50_55_eps_0_1}. Note that the difference between the amplitude envelopes of the symmetric mode and the anti-symmetric mode decreases with an increase in beam lengths.

We now proceed to demonstrate that a continuous 1-parameter family of solutions corresponding to arbritrarily phase shifted deformations of the beam can be generated for extremely long lengths (cf. Section \ref{sec:Asymp_Analysis}) using the finite-element model and a beam length of $L=L_a$. For this we require longer lengths of the beam than the four cases considered above. We explain this at the end of this section.

We note that the asymptotic amplitude function \eqref{eq:Asymptotic_ASymm_Amplitude} takes a maximum value at $X=0$ for $X\in[- \varepsilon L_{a}/2, \varepsilon L_{a}/2]$ and decreases towards $X=\pm \varepsilon L_a/2$. Furthermore, we see from \eqref{eq:Difference_Amplitude} that $w_{a,\phi}$ misses the zero-displacement boundary conditions at most by $\tilde{w}:=\max_{\phi\in[0,2\pi)} w_{a,\phi}(\pm L_{a}/2)=\varepsilon A^{*}(\varepsilon L_{a}/2)$. We choose $L_a=m\pi$ large enough so that $\tilde{w}\approx0$, say, for $\varepsilon=0.1$. In particular, $m=200$ gives $\tilde{w}=\mathcal{O}(10^{-8})$, ensuring that the simply-supported boundary conditions are satisfied for any phase angle $\phi\in[0,2\pi)$.

\begin{figure}[H]
    \centering
        \includegraphics[width=0.85\textwidth]{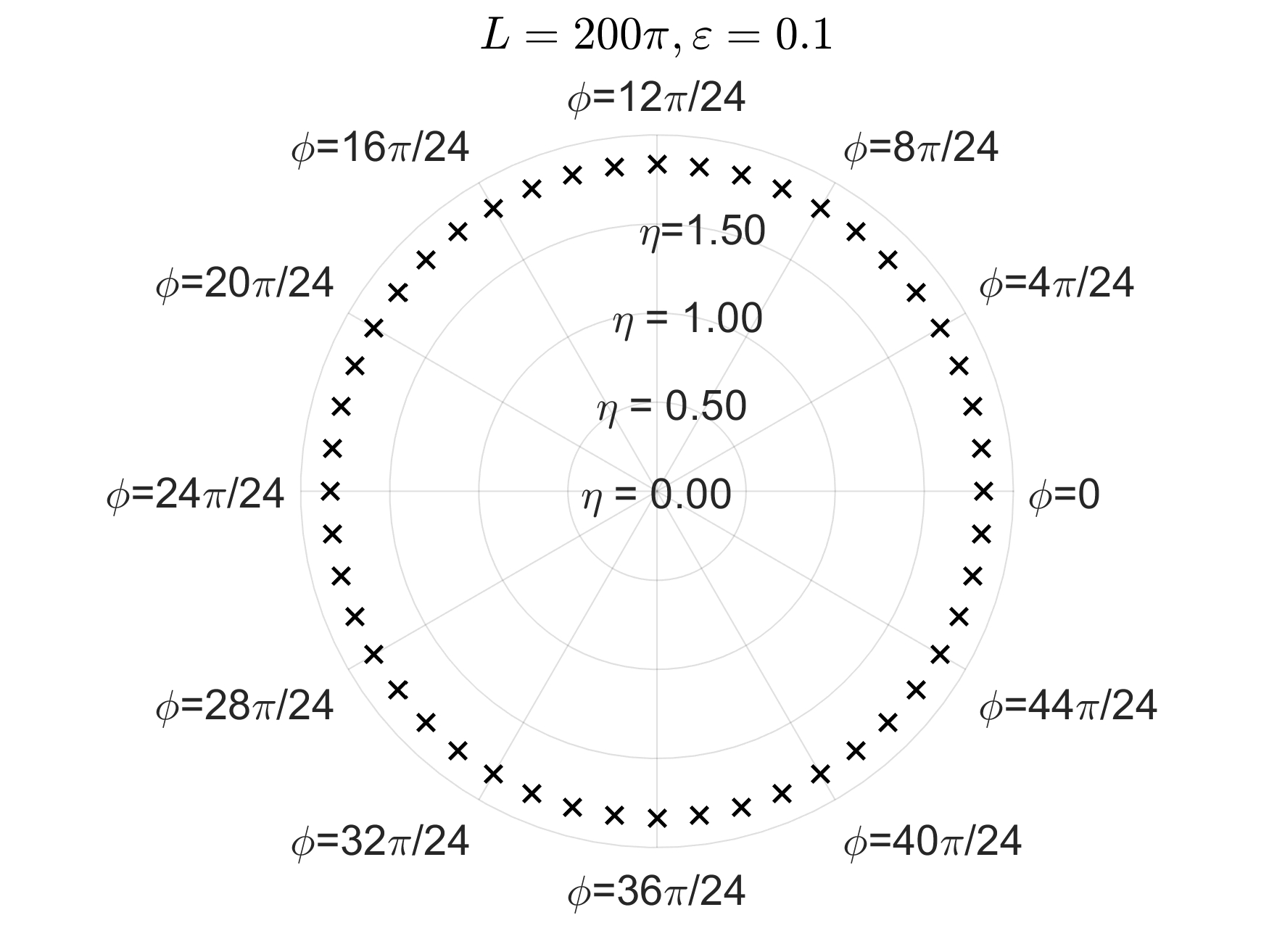}
   \caption{$\eta=\Vert w_{a,\phi}\Vert_{2}$ vs. $\phi=$ phase-angle}
   \label{fig:Polarplot_eta_phi_eps_0_1}
\end{figure}

The numerically computed beam deformations are obtained by employing the asymptotic displacement function $w_{a,\phi}$ given in \eqref{eq:Difference_Amplitude} as the initial trial solution for the discretized model at $L_a=200\pi$ and $\varepsilon=0.1$. Newton's method is then used iteratively. For any chosen $\phi \in [0, 2\pi)$ the implementation converges readily. Figure \ref{fig:Polarplot_eta_phi_eps_0_1} gives a polar plot of the $\ell^{2}$- norm as a function of $\phi=\phi(n):=n \pi/24, n=0, 1,\ldots 47$, for the numerically computed solution points which align exactly on a circle. Symmetric solutions correspond to $n=12, 36$, anti-symmetric solutions correspond to $n=0, 24$, and all the other values of $n$ correspond to solutions that are neither even nor odd. 

\begin{figure}[H]
    \centering
    \begin{subfigure}[H]{0.80\textwidth}
        \centering
        \includegraphics[width=0.90\textwidth]{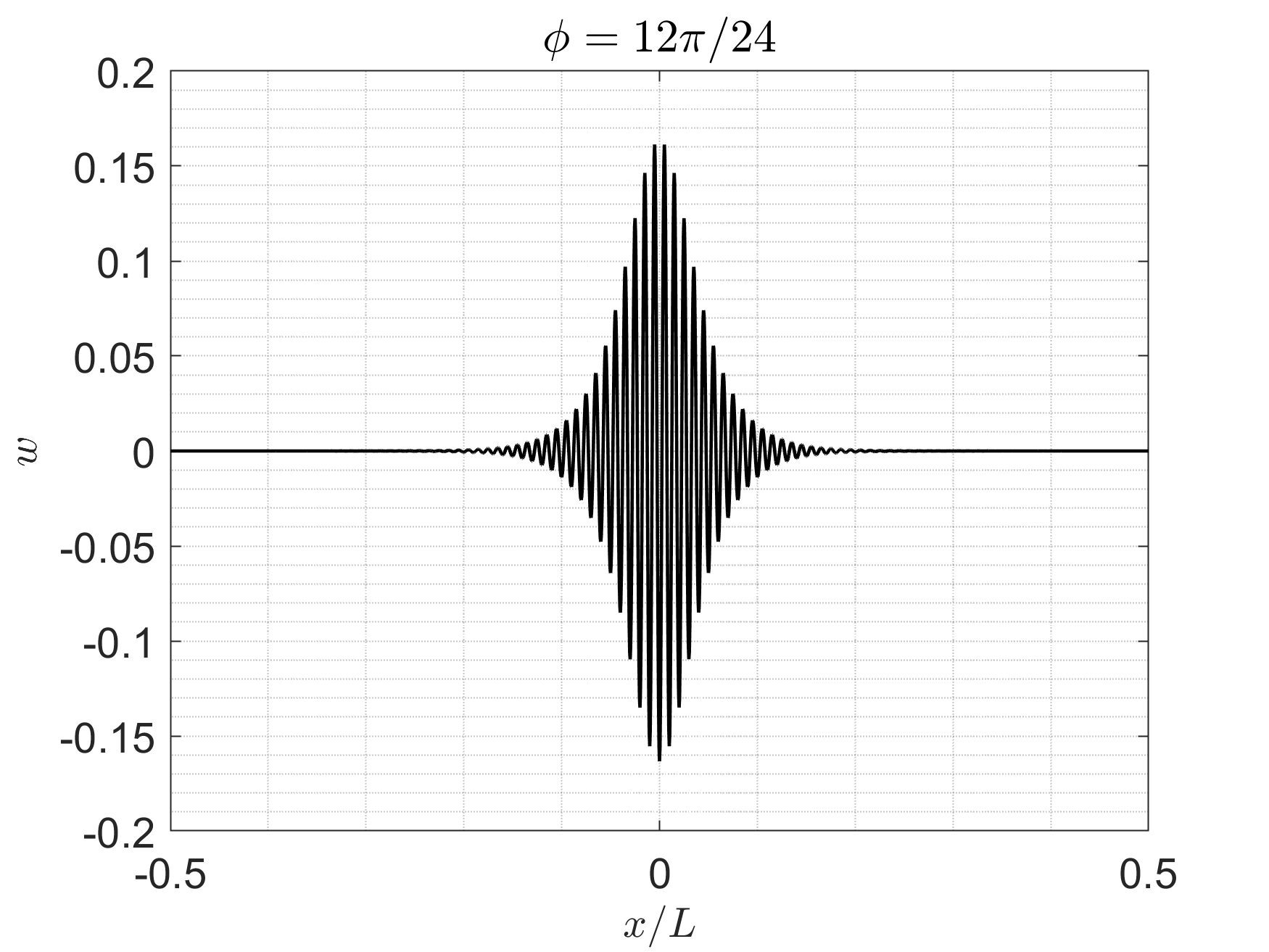}
        \caption{}    
    \end{subfigure}
    \hfill 
    \begin{subfigure}[H]{0.80\textwidth}
        \centering
        \includegraphics[width=0.90\textwidth]{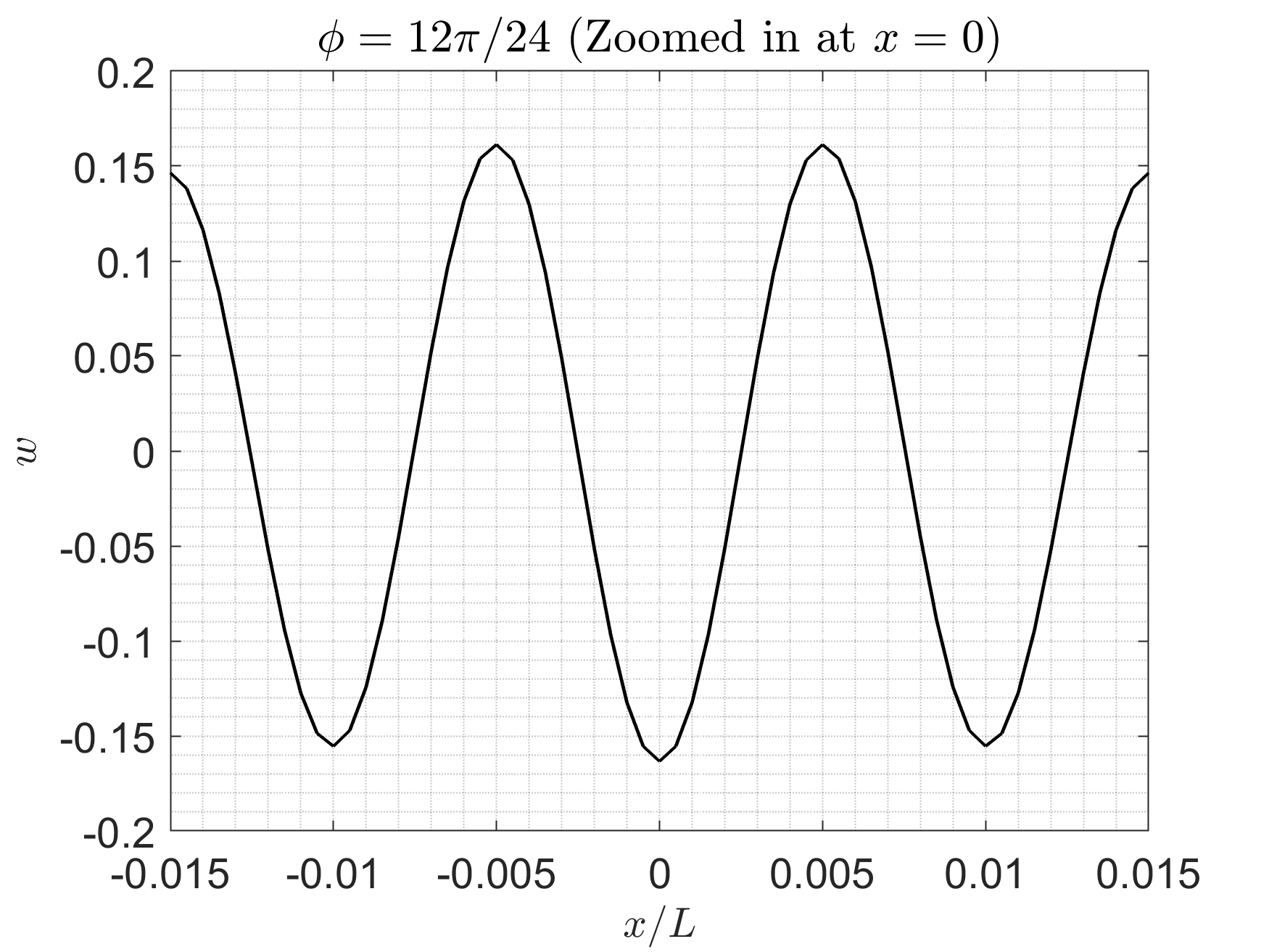}
        \caption{}   
    \end{subfigure}
\caption{Symmetric}    
\label{fig:phi_12pi_24_eps_0_1}
\end{figure}

\begin{figure}[H]
    \centering
    \begin{subfigure}[H]{0.80\textwidth}
        \centering
        \includegraphics[width=0.90\textwidth]{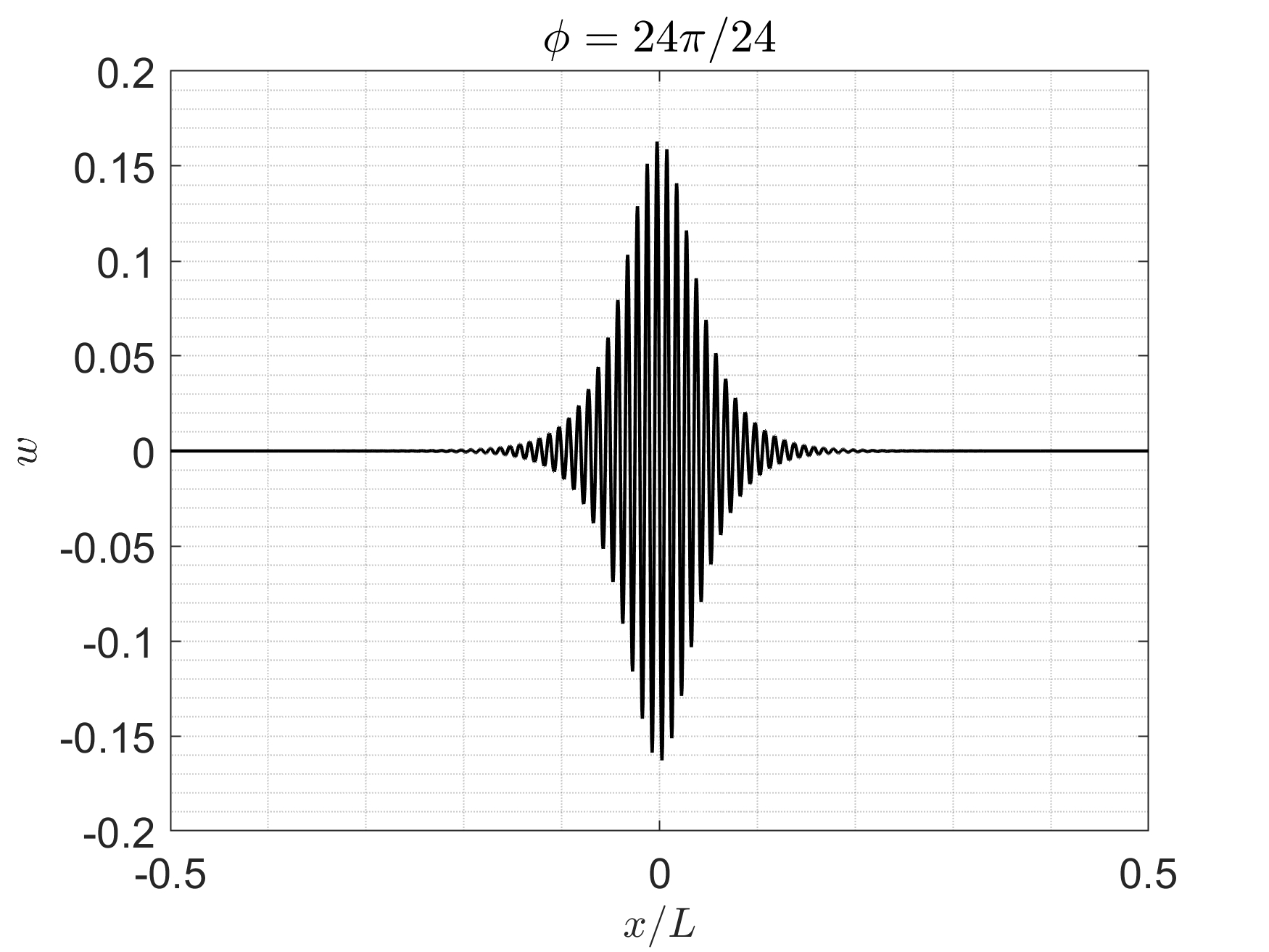}
        \caption{}    
    \end{subfigure}
    \hfill 
    \begin{subfigure}[H]{0.80\textwidth}
        \centering
        \includegraphics[width=0.90\textwidth]{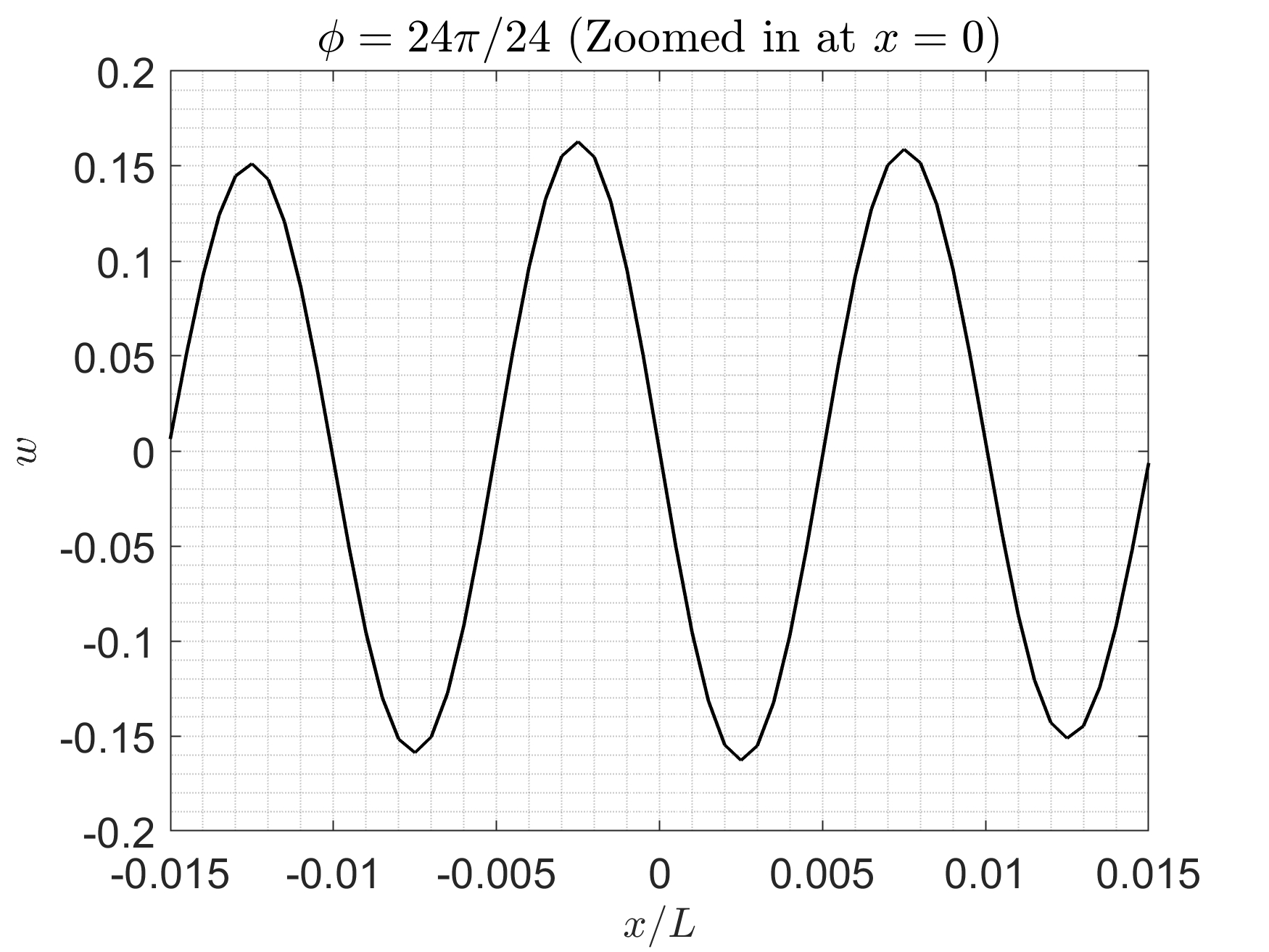}
        \caption{}   
    \end{subfigure}
\caption{Anti-symmetric}    
\label{fig:phi_24pi_24_eps_0_1}
\end{figure}

\begin{figure}[H]
    \centering
    \begin{subfigure}[H]{0.80\textwidth}
        \centering
        \includegraphics[width=0.90\textwidth]{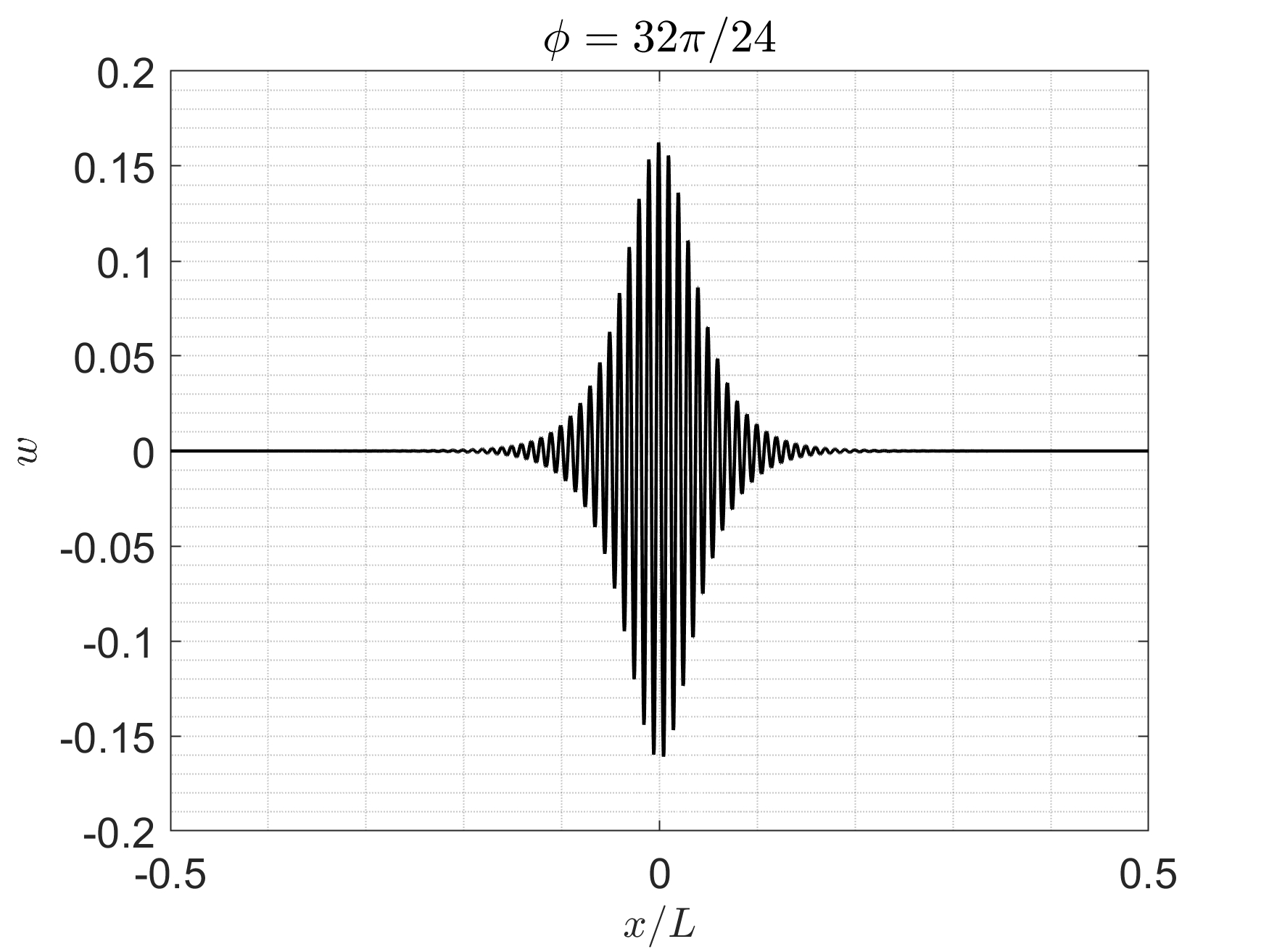}
        \caption{}    
    \end{subfigure}
    \hfill 
    \begin{subfigure}[H]{0.80\textwidth}
        \centering
        \includegraphics[width=0.90\textwidth]{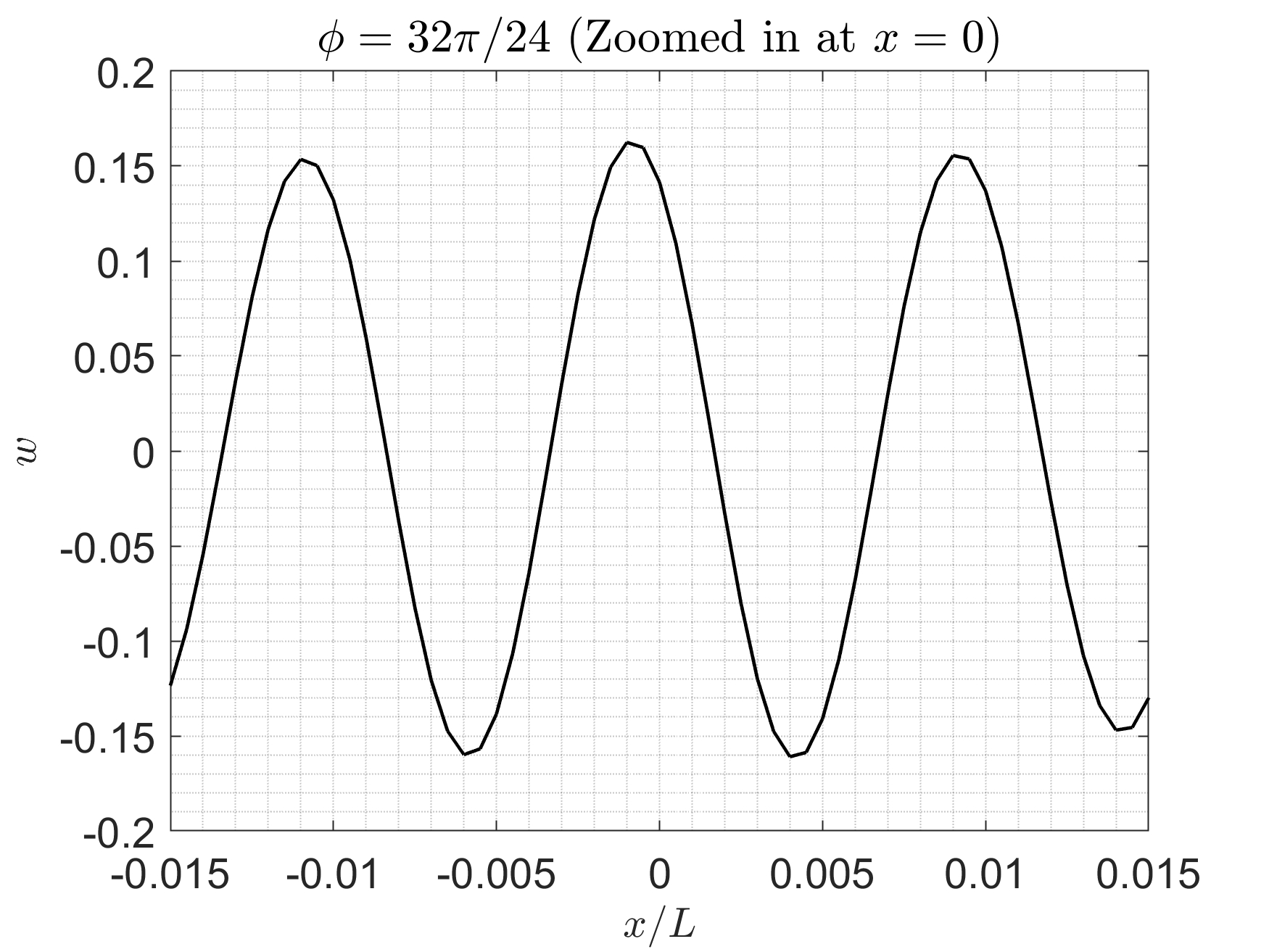}
        \caption{}   
    \end{subfigure}
\caption{Neither Symmetric nor Anti-symmetric}    
\label{fig:phi_32pi_24_eps_0_1}
\end{figure}

Figures \ref{fig:phi_12pi_24_eps_0_1}-\ref{fig:phi_32pi_24_eps_0_1} give a sampling of the numerical solutions of the beam displacement field obtained as described above.

For the four lengths considered earlier, cf. Table \ref{table:Secondary_Bifurcation_points}, the solutions do not converge with the procedure mentioned above for any arbitrary $\phi$ as $\tilde{w}=\mathcal{O}(10^{-3})$ fails to accurately account for zero-displacement boundary conditions. 

\section{Conclusion}\label{sec:Section_6}

As mentioned in the Introduction, this work is motivated by the symmetry properties of the solution for wrinkling in highly stretched thin elastic membranes\cite{Healey_Li_Cheng}. A direct comparison of our analysis here for the ``long-beam" problem with the more complex wrinkling problem is not possible. Nonetheless, both models exhibit the same phenomenon, viz., behavior associated with a continuous symmetry group for a problem nominally possessing a finite symmetry group. In the beam problem, the latter corresponds to simple even-odd symmetry with respect to the origin. Yet, as the non-dimensional length $L$ becomes sufficiently large, an additional translation symmetry emerges asymptotically. Note from \eqref{eq:TPE} that $L$ sufficiently large can be associated with the beam thickness $h$ sufficiently small (since $I = b h^3$) and/or with sufficiently large foundation stiffness $k_2$. Although the compressive-stress distribution in the wrinkling problem is not uniform, an analogy with the beam problem is clear. The transverse width $W$ of the membrane is analogous to the actual length $\bar{L}$ of the beam. The characteristic width $W_c$ is then chosen in precisely the same way, where $h$ is now the membrane thickness (with $b = 1$). Moreover, as pointed out in \cite{Healey_Li_Cheng}, an equivalent linear foundation stiffness $k_2$ can be associated with the tensile stress (per unit length) in the highly stretched direction of the membrane. The thickness $h$ of the membrane is extremely small, while the tensile stress is extremely large, cf. \cite{Healey_Li_Cheng}. Consequently, the effective width of the membrane is enormous.

Our results demonstrate the asymptotic emergence of an orbit of solutions in a problem having only a finite complete symmetry group.  We mention that in the analysis of a closely related beam problem, an asymptotic correction to the exact infinite-length solution for sufficiently long finite beams is obtained in \cite{Rivetti_Neukirch}. The construction there is based on an ansatz that is restricted to either symmetric or anti-symmetric solutions. In particular, asymptotic orbits of solutions are not addressed in that work. \\
\\
{\bf{Acknowledgments}}: This work was supported in part by grants from \'Ecole Polytechnique and C.N.R.S. (Centre National de Recherche Scientifique) during the AY 2017-2018, while TJH was a Distinguished Visiting Professor at the Laboratoire de M\'ecanique des Solides. The work of SSP and TJH was also supported in part by the National Science Foundation through grant DMS-1613753.

\end{document}